\documentclass[aps,nofootinbib,twocolumn,superscriptaddress]{revtex4-1}
 
\usepackage{amsmath,amssymb,euscript} 
\usepackage{graphicx}  
\usepackage{braket}
\usepackage{bbm} 
\usepackage{color}

\def\dd{d}
\def\ii{i}
\def\e{e}

\RequirePackage[
   hyperindex,colorlinks,bookmarksnumbered,
   plainpages=true,pdfstartview=FitH]{hyperref}
\hypersetup{linkcolor=blue,urlcolor=blue,citecolor=blue}
\usepackage{hyperref}
\usepackage{breakurl}

\usepackage{ulem} 
\renewcommand{\emph}[1]{\textit{#1}}
\definecolor{purple}{rgb}{0.5,0,0.6}
\definecolor{darkblue}{rgb}{0,0,0.5}
\definecolor{darkgreen}{rgb}{0,0.5,0}
\definecolor{darkred}{rgb}{.7,0,0}
\definecolor{purple}{rgb}{0.5,0,0.6}
\definecolor{orange}{rgb}{1,0.5,0}
\definecolor{grey}{rgb}{.6,.6,.6}
\definecolor{lightpink}{rgb}{1,0.7,0.75}
\definecolor{pink}{rgb}{1,0.4,0.58}
\definecolor{deeppink}{rgb}{1,0.08,0.58}

\newcommand{\todo}[1]{}

\newcommand{\pdag}{{\phantom{\dagger}}}
\newcommand{\pstar}{{\phantom{\star}}}

\renewcommand{\emph}[1]{\textit{#1}} 

\newcommand{\Eq}[1]{Eq.\,(\ref{#1})}

\newcommand{\Lorentz}[2]{\delta_{{#1}}(\omega-\varepsilon_{#2})}

\newcommand{\rhotilde}{\varrho}
\def\C{C}

\begin{document}
\title{Lindblad-Driven Discretized Leads for Non-Equilibrium 
Steady-State Transport in Quantum Impurity Models: Recovering the Continuum Limit}
\date{\today}
\author{F.~Schwarz}\affiliation{Physics  Department,  Arnold  Sommerfeld  Center  for  Theoretical  Physics,  and  Center  for  NanoScience,
Ludwig-Maximilians-Universit\"at,  Theresienstra{\ss}e  37,  80333  M\"unchen,  Germany}
\author{M.~Goldstein}\affiliation{Raymond and Beverly Sackler School of Physics and Astronomy, Tel Aviv University, Tel Aviv 6997801, Israel}
\author{A.~Dorda}\affiliation{Institute of Theoretical and Computational Physics, 
Graz University of Technology, 8010 Graz, Austria}
\author{E.~Arrigoni}\affiliation{Institute of Theoretical and Computational Physics,
Graz University of Technology, 8010 Graz, Austria}
\author{A.~Weichselbaum}\affiliation{Physics  Department,  Arnold  Sommerfeld  Center  for  Theoretical  Physics,  and  Center  for  NanoScience,
Ludwig-Maximilians-Universit\"at,  Theresienstra{\ss}e  37,  80333  M\"unchen,  Germany}
\author{J.~von Delft}\affiliation{Physics  Department,  Arnold  Sommerfeld  Center  for  Theoretical  Physics,  and  Center  for  NanoScience,
Ludwig-Maximilians-Universit\"at,  Theresienstra{\ss}e  37,  80333  M\"unchen,  Germany}

\begin{abstract}
The description of interacting quantum impurity models in steady-state nonequilibrium is an open challenge for computational many-particle methods: the numerical requirement of using a finite number of lead levels and the physical requirement of describing a truly open quantum system are seemingly incompatible. One possibility to bridge this gap is the use of Lindblad-driven discretized leads (LDDL): one couples auxiliary continuous reservoirs to the discretized lead levels and represents these additional reservoirs by Lindblad terms in the Liouville equation. For quadratic models governed by Lindbladian dynamics, we present an elementary approach for obtaining correlation functions analytically. In a second part, we use this approach to explicitly discuss the conditions under which the continuum limit of the LDDL approach recovers the correct representation of thermal reservoirs. As an analytically solvable example, the nonequilibrium resonant level model is studied in greater detail. Lastly, we present ideas towards a numerical evaluation of the suggested Lindblad equation for interacting impurities based on matrix product states. In particular, we present a reformulation of the Lindblad equation, which has the useful property that the leads can be mapped onto a chain where both the Hamiltonian dynamics and the Lindblad driving are local at the same time. Moreover, we discuss the possibility to combine the Lindblad approach with a logarithmic discretization needed for the exploration of exponentially small energy scales.
\end{abstract}

\maketitle

\section{Introduction}
Quantum impurity models describe discrete local quantum degrees of freedom coupled to continuous baths of excitations. They were originally introduced for the description of magnetic impurities in metals, but in the last two decades became highly relevant also for describing transport through quantum dots or nanotubes coupled to metallic leads. While some notable impurity models are integrable, others are not; hence our interest here will be directed towards nonperturbative numerical many-body methods. In experimental work on such systems it is routine to measure the nonlinear current-voltage characteristics. However, numerically calculating such steady-state nonequilibrium properties is a difficult computational problem that is by no means routine. Despite much effort and noteworthy progress for some benchmark problems such as the interacting resonant level model, the Kondo model and the single-level Anderson impurity model \cite{PhysRevLett.101.140601,1367-2630-12-4-043042, PhysRevB.79.235336,PhysRevLett.101.066804, PhysRevB.90.085121,PhysRevLett.112.146802,PhysRevLett.116.036801}, the theoretical description of steady-state nonequilibrium can still be regarded as a major open challenge for computational treatments of quantum impurity models.
 
The two key ingredients, local interactions and steady-state transport, in computational practice lead to a set of requirements that are hard to reconcile. (i) The presence of \textit{interactions} means that the models of interest are not quadratic; hence their treatment requires many-body methods.  (ii) These methods should be able to reach very \textit{low energy scales} since quantum impurity models often show interesting many-body correlations below a characteristic, exponentially small low-energy scale (e.g.,\ the Kondo temperature for the Kondo or Anderson models). (iii) Steady-state transport means that charge flows at a constant rate in at one side and out on the other. Describing this properly requires dealing with a truly \textit{open} quantum system.

For equilibrium situations, where (iii) is not relevant, two powerful approaches based on matrix product states (MPS) are available, which both use a discretized description of the leads, formulated in terms of one-dimensional chains of finite length. The first is Wilson's numerical renormalization group (NRG) \cite{RevModPhys.47.773,RevModPhys.80.395}. It fulfils the requirement (i) as it is based on an iterative diagonalization of the full many-body Hamiltonian, and it complies with condition (ii) by discretizing the leads on a logarithmic grid capable of resolving exponentially small energy scales. The second method is the density matrix renormalization group (DMRG), which can be used also in situations where a logarithmic discretization is not advisable, albeit at the cost of requiring very long chains to resolve small energy scales. However, both these approaches treat the impurity plus discretized leads as a truly \textit{closed} quantum system and, hence, are fundamentally limited in dealing with the open-system requirement (iii) \cite{Rosch2012}. Although there are ideas on how to extend the use of NRG to situations of steady-state nonequilibrium \cite{PhysRevLett.101.066804} and although progress has been made using time-dependent DMRG (tDMRG) approaches \cite{PhysRevLett.101.140601, 1367-2630-12-4-043042,PhysRevB.79.235336}, it would be highly desirable to have a versatile strategy based on MPS methodology that intrinsically overcomes the discrepancy between the numerical need to discretize the leads on the one hand and the requirement of a truly open quantum system on the other hand.

During the last few years, a new scheme has been put forth \cite{1.3658736,1.3548065,1.3581098,0953-8984-24-22-225304,1751-8121-47-9-095002,1751-8121-48-1-015004,PhysRevB.86.125111,1402-4896-86-5-058501,PhysRevB.87.195114,PhysRevLett.110.086403,PhysRevB.89.165105,PhysRevB.92.125145} to address this discrepancy. Its main idea is to introduce additional continuous reservoirs coupled to the discretized leads to render the system truly open again. Since these additional reservoirs are then described using Lindblad operators, we will call the approach \textit{Lindblad-driven discretized leads} (LDDL).  Consider an arbitrary impurity and noninteracting leads enumerated by a lead index $\alpha$. In the thermodynamic limit, lead $\alpha$ is continuous in energy. This may be coarse-grained in energy using discrete levels $q$, such that each level $q$ now represents an entire energy interval. The continuum limit will be recovered if each level $q$ is coupled to the remainder of the states in the energy interval it represents, which thus serves as an environmental reservoir for it. Now, for the description of steady-state nonequilibrium physics, one has to ensure that each discretized lead $\alpha$ is held at a fixed temperature $T_\alpha$ and at a fixed chemical potential $\mu_\alpha$. In the LDDL scheme, this is achieved by embedding the system $S$ consisting of impurity and discretized lead levels~$q$ into an environment $R$. This environment consists of one reservoir $\mathcal{R}_q$ for each discrete lead level~$q$, to be associated with the above-mentioned continuum of levels which that level represents, and is described by Lindblad driving terms in the Liouville equation for the density matrix of the subsystem $S$. The driving rates involved in these Lindblad terms have to be chosen such that the occupation numbers for the lead levels are driven towards the values that they would have if the leads were decoupled from the impurity, namely $f_\alpha(\varepsilon_q)$, where $f_\alpha(\omega)$ is the Fermi distribution characterizing lead $\alpha$, and $\varepsilon_q$ the energy associated with lead level~$q$.

The initial publications utilizing the LDDL scheme presented various pieces of evidence that it offers a viable way for describing nonequilibrium steady-state transport in quantum impurity models. References~[\onlinecite{1.3548065,1.3581098,0953-8984-24-22-225304,1751-8121-47-9-095002,1751-8121-48-1-015004}] used it as a starting point for analytical methods like perturbative and mean-field approaches or the coupled cluster method in superoperator representations. 
In these models, the driving rates occurring in the Lindblad equation were viewed as phenomenological parameters, and we adopt the same point of view here. We note, though, that it should be possible to formally derive these driving rates using the reaction coordinate method \cite{1.3490188,1.3532408,1.4866769}.
In Refs.~[\onlinecite{PhysRevB.86.125111,1402-4896-86-5-058501,PhysRevB.87.195114}] the LDDL Lindblad equation was evaluated based on a method established in Ref.~[\onlinecite{Prosen2010}]. More recently, Refs.~[\onlinecite{PhysRevLett.110.086403,PhysRevB.89.165105,PhysRevB.92.125145}] presented an alternative version of the LDDL approach based on a fit procedure for the Lindblad coefficients. Ideas similar to the LDDL approach have also been applied in the context of spin transport in quantum chains \cite{1742-5468-2009-02-P02035,373001,PhysRevLett.107.060403,025016}. Furthermore, in close relation to the LDDL scheme, Refs.~[\onlinecite{PhysRevLett.111.040601,1367-2630-18-8-083006}] suggest the use of discrete modes coupled to a continuum bath to explore analogues of quantum transport in experimental devices that actually have a reduced number of degrees of freedom.

The LDDL approach relies on a decomposition of the bath into a discrete part coupled to the impurity in which many-body effects can be considered, and a continuous remainder which reduces finite-size effects. The same idea also forms the basis  of the embedded-cluster approximation \cite{PhysRevLett.82.5088,0953-8984-15-50-013,PhysRevB.78.085308}.

Our own long-term interests lie in using the LDDL scheme as starting point for numerical computations that seek to solve the Liouville equation for the many-body density matrix of the system $S$ using MPS methods. Compared to standard equilibrium calculations, where one deals with many-body quantum states, solving the Liouville equation would involve calculating many-body density matrices, and hence be computationally more demanding. Nevertheless, we believe this to be worth the additional effort, because of the direct, explicit way in which the LDDL scheme addresses the open-system requirement (iii).  Moreover, there has been much recent progress in MPS-based approaches for solving Liouville equations describing open quantum systems \cite{1742-5468-2009-02-P02035,Daley2014,PhysRevLett.116.237201,PhysRevLett.114.220601,PhysRevA.92.022116,PhysRevLett.102.040402,PhysRevA.82.063605,PhysRevLett.103.240401,PhysRevA.84.041606}, some of which seem directly suitable for tackling the Lindblad equation arising in the LDDL scheme. In particular, already in 2009, transport in spin chains was described using a matrix product operator (MPO) ansatz combined with Lindblad reservoirs \cite{1742-5468-2009-02-P02035}. More recently, an LDDL scheme together with MPOs was used to investigate the nonequilibrium properties of an Anderson impurity \cite{PhysRevB.92.125145}.

In the present paper, which is intended to set the stage for such future MPS-based works, we address three preliminary but important general questions. (i) How should the Lindblad rates in the LDDL scheme be chosen in order to properly recover the continuum limit?  (ii) Is it possible to formulate the Lindblad driving terms in such a way that they remain local when the leads are mapped to chains with local Hamiltonian dynamics?  (iii) Can the LDDL scheme be used in conjunction with the logarithmic discretization of lead states needed for the exploration of exponentially small energy scales? Questions (i) and (ii) can actually be addressed fully in the context of purely \textit{non}-interacting quantum impurity models. The reason is that for any quantum impurity model, with or without local interactions, the lead properties needed to specify the steady-state dynamics are \textit{fully} encoded in the \textit{bare} (i.e.,\ with zero lead-impurity coupling) steady-state correlators of that linear combination of lead operators that couples to the impurity. 

To answer question (i), it suffices to identify the Lindblad driving conditions that reproduce the bare steady-state correlators known for continuum leads. Our main conclusion in this regard is, perhaps not surprisingly, that the broadening of the discretized levels generated by the Lindblad driving should be such that the resulting level width for each level is comparable to the level spacing to neighboring lead levels. This result is consistent with the conclusions of previous works utilizing the LDDL scheme, in particular in Ref.~[\onlinecite{1.3548065}], which also addressed the question of how to recover the continuum limit. Questions (ii) and (iii) have not received much attention previously. We conclude that both can be answered affirmatively, thus opening the door towards treating LDDL systems using MPS-based methods in the near future. 
  
The rest of this paper is organized as follows: first, considering a completely generic quadratic Lindblad equation (Sec.~\ref{sec: Greens functions}), we present a simple derivation of analytical formulas for the system's steady-state correlators. This reproduces results found previously using rather more elaborate methods involving superoperators \cite{PhysRevLett.110.086403,PhysRevB.89.165105}. The derivation offered here is so elementary that we believe it to be of general interest (also beyond the context of quantum impurity models).  Second, we use these results to obtain analytical expressions for the steady-state lead correlators. These allow us to identify the choice of Lindblad parameters that ensures that the leads within the LDDL scheme become equivalent to thermal reservoirs in the continuum limit, thus answering question (i) (Sec.~\ref{sec: Lindblad equation for impurity models} and \ref{sec: hybridization}). As an explicit example of a non-interacting impurity model, where the full Liouville equation can be solved analytically, we study the nonequilibrium resonant level model (RLM) in some detail (Sec.~\ref{sec: Greens functions RLM} and \ref{sec: current RLM}). The results obtained by our elementary treatment are consistent with the ones obtained previously for this model using the superoperator formalism \cite{1.3548065} and instructively illustrate under what conditions the continuum limit is recovered. Sections~\ref{sec: Local chain}~and~\ref{sec: Logarithmic Discretization} are devoted to questions (ii) and (iii) regarding local Lindblad driving and logarithmic discretization, respectively. Section~\ref{sec:conclusion-outlook} summarizes our conclusions. Finally, Appendix~\ref{sec: appendix A} discusses some details arising in the context of logarithmic discretization, and in Appendix~\ref{sec: appendix B}, a fermionic version of the quantum regression theorem is derived.


\section{Green's functions in the Lindblad approach}\label{sec: Greens functions}
In this section we introduce Green's functions for systems that evolve in time under Lindbladian dynamics. For quadratic systems we derive closed expressions for the steady-state Green's functions. This section, therefore, is not restricted to impurity models, but the formulas derived for quadratic models lay the foundation for an analytical exploration of the LDDL scheme presented in Secs.~\ref{sec: Lindblad approach to impurity systems}-\ref{sec: Logarithmic Discretization}.

\subsection{The Lindblad equation}
Consider a system~$S$ linearly coupled to a large reservoir~$R$ which together form a closed quantum system with Hamiltonian dynamics described by the full Hamiltonian of system and reservoir, $H_\text{full}$. Equal-time expectation values are defined by
\begin{align}
\label{eq: equal time full}\braket{A(t)}&=\text{tr}_{S,R}\left(A(t)\,\rho_\text{full}\right)=\text{tr}_{S,R}\left(A\,\rho_\text{full}(t)\right)\,,
\end{align}
where $A$ acts on the system $S$, and the time evolution of $A(t)$ and of the full density matrix $\rho_\text{full}(t)$ is given by (with \mbox{$\hbar=1$})
\begin{subequations}
\begin{align}
A(t)&=\e^{\ii H_\text{full}t}\,A\,\e^{-\ii H_\text{full}t}\,,\\
\rho_\text{full}(t)&=\e^{-\ii H_\text{full}t}\,\rho_\text{full}\,\e^{\ii H_\text{full}t}\,.
\end{align}
\end{subequations}
Two-point correlators for operators $A$ and $\C$ acting on $S$ are defined as
\begin{subequations}\label{eq: correlators full}
\begin{align}
\braket{A(t)\C}&=\text{tr}_{S,R}\left(A(t)\C\,\rho_\text{full}\right)=\text{tr}_{S,R}\left(A\,{\rhotilde}_{\C,\text{full}}(t)\right)\,,\\
\braket{\C A(t)}&=\text{tr}_{S,R}\left(A(t)\rho_\text{full} \C\right)=\text{tr}_{S,R}\left(A\,{\rhotilde}'_{\C,\text{full}}(t)\right)\,,
\end{align}
\end{subequations}
where the $\C$-dependent auxiliary operators ${\rhotilde}_{\C,\text{full}}(t)$ and ${\rhotilde}_{\C,\text{full}}'(t)$ are defined by
\begin{subequations}\label{eq: tilde rho def and Hamiltonian dynamics}
\begin{align}
\label{eq: tilde rho def}{\rhotilde}_{\C,\text{full}}(t=0)&=\C\rho_\text{full}\,,\quad {\rhotilde}'_{\C,\text{full}}(t=0)=\rho_\text{full}\C\,,\\
\label{eq: tilde rho Hamiltonian dynamics}{\rhotilde}_{\C,\text{full}}^{(\prime)}(t)&=\e^{-\ii H_\text{full}t}\,{\rhotilde}_{\C,\text{full}}^{(\prime)}\,\e^{\ii H_\text{full}t}\,.
\end{align}
\end{subequations}

If the reservoir $R$ is Markovian, its degrees of freedom can be traced out using quite general assumptions~\cite{Breuer2002}. The resulting equation for the time evolution of the reduced density matrix of system $S$, $\rho(t)=\text{tr}_R(\rho_\text{full}(t))$, known as Lindblad equation \cite{CommMathPhys48.119,CommMathPhys65.281}, can always be written in the form~\cite{Gardiner2000, Breuer2002} 
\begin{subequations}\label{eq: Lindblad general}
\begin{align}
\dot{\rho}(t)&=\mathcal{L}\rho(t)=-\ii\left[H,\rho(t)\right]+\mathcal{D}\rho(t)\,,\\
\mathcal{D}\rho(t)&=\sum_{m}\left(2J_{m}\rho(t)J_{m}^\dagger-\left\{J_{m}^\dagger J_{m},\rho(t)\right\}\right)\,.
\end{align}
\end{subequations}
The unitary operator~$H$ describes the Hamiltonian part of the dynamics. It is not necessarily equal to that part of the original full Hamiltonian that acts on system~$S$, but can contain additional Lamb shifts [cf. Eq.\,(\ref{eq: Lamb shift}) below]. $\displaystyle{\mathcal{D}\rho(t)}$ describes the dissipative part of the time evolution. The so-called Lindblad operators~$J$ act on system~$S$ and are unconstrained otherwise, e.g., are not normalized. Note that the Lindblad equation is only valid for $t>0$\,. By construction, it preserves the positivity and the trace of the density matrix.

\subsection{Steady-state Green's functions for quadratic models}
For a system with quadratic Hamiltonian governed by Lindbladian dynamics with linear Lindblad operators, it is possible to find closed expressions for steady-state correlation functions, see Eqs.\ (\ref{eq: GR}) and (\ref{eq: GK}) below. For example, in Refs.~[\onlinecite{PhysRevLett.110.086403,PhysRevB.89.165105}], they were derived using superoperators. Here, we offer a simple complementary derivation which utilizes only elementary definitions.

Our starting point is a quadratic system $S$ coupled linearly to a quadratic reservoir $R$. We write the Hamiltonian of system $S$ as 
\begin{flalign}
\label{eq:quadratic-Hamiltonian}
H&=\sum_{mn}h_{mn}L_m^\dagger L_n \,, 
\end{flalign}
with $\left\{L_m,L_n^\dagger\right\}=\delta_{mn}$, $\left\{L_m,L_n\right\}=0$. The operators $L_m^{(\dagger)}$ will act as normalized Lindblad operators later on. Furthermore, in contrast to the operators $J_m^{(\dagger)}$ in Eq.\,(\ref{eq: Lindblad general}), we now distinguish explicitly between annihilation ($L_m$) and creation operators ($L_m^\dagger$). To fully characterize the system's nonequilibrium steady-state (NESS) physics, we will be interested in the retarded, advanced and Keldysh Green's functions of $S$ in the steady state \cite{Haug1997,Kamenev2011}, ${\mathcal G}^{R/A/K}(t)$, and their Fourier transforms, $G^{R/A/K}(\omega)$, defined as follows: 
\begin{subequations}
\begin{align}
\label{eq: GR definition}\mathcal{G}_{mn}^{R}(t)=&-\ii\,\theta(t)\braket{\left\{L_m(t),L_n^\dagger\right\}}_\text{NESS}\,,\\
\label{eq: GA definition}\mathcal{G}_{mn}^{A}(t)=&\,\ii\,\theta(-t)\braket{\left\{L_m(t),L_n^\dagger\right\}}_\text{NESS}\,,\\
\label{eq: GK definition}\mathcal{G}^K_{mn}(t)=&-\ii\braket{\left[L_m(t),L_n^\dagger\right]}_\text{NESS}\,,\\
\label{eq: G Fourier} G_{mn}^{R/A/K}(\omega)=&\int_{-\infty}^{\infty}\dd t\,\e^{\ii\omega t}\mathcal{G}_{mn}^{R/A/K}(t)\,, 
\end{align}
\end{subequations}
with $\theta(t)$ the Heaviside step function.
Since the steady state is translationally invariant in time, these Green's functions satisfy the relations
\begin{align}
\label{eq: time invariance relations}G^{R/A}(\omega) = G^{A/R \dagger} (\omega) , \quad {\mathcal G}^K(t) = - {\mathcal G}^{K\dagger} (-t)\,, 
\end{align}
where matrix notation is understood.

Formally, these correlators can be evaluated by integrating out the reservoir $R$, leading to the following expressions:
\begin{subequations}
\begin{align}
\label{eq: GR Keldysh}G^R_\text{exact}(\omega)&=\left(\omega-h-\Sigma_\text{exact}^R(\omega)\right)^{-1}\,,\\
\label{eq: GK Keldysh}G^K_\text{exact}(\omega)&=G_\text{exact}^R(\omega)\Sigma_\text{exact}^K(\omega)G_\text{exact}^A(\omega)
\end{align}
\end{subequations}
These express the effect of $R$ on $S$ fully in terms of the retarded and Keldysh component of the self-energy $\Sigma_{\rm exact}^{R/K}(\omega)$, in which all information about the reservoir is encoded. While for interacting systems the self-energy will contain additional terms due to the interaction, for quadratic systems $\Sigma_\text{exact}^{R/K}(\omega)$ simply describes the hybridization between system $S$ and reservoir $R$ and can therefore be calculated explicitly.

Here, we are interested in the less complete description that results from making Markovian approximations in treating the reservoir and encoding its effects only at the level of a Liouville equation for the system density matrix~$\rho$. For a fully quadratic system, the most general form of the resulting Lindblad equation is 
\begin{align}\label{eq: Lindblad quadratic}
\notag\dot{\rho}(t)&=-\ii[\tilde{H},\rho(t)]\\
\notag+&\sum_{mn}\Lambda_{mn}^{(1)}\left(2L_{n}^\pdag\rho(t)L_{m}^\dagger-\left\{L_{m}^\dagger L_{n}^\pdag,\rho(t)\right\}\right)\\
+&\sum_{mn}\Lambda_{mn}^{(2)}\left(2L_{m}^\dagger\rho(t)L_{n}^\pdag-\left\{L_{n}^\pdag L_{m}^\dagger,\rho(t)\right\}\right)\,,
\intertext{where the matrices $\Lambda^{(1,2)}$ are Hermitian and positive. The effective Hamiltonian of the system,}
\label{eq: Lamb shift}\tilde{H}&=\sum_{mn}\tilde{h}_{mn}L_m^\dagger L_n^\pdag=\sum_{mn}\left(h_{mn}+\Delta^{\rm{Lamb}}_{mn}\right)L_m^\dagger L_n^\pdag
\end{align}
contains the Lamb shift $\Delta^{\rm{Lamb}}$ corresponding to an effective shift of the energies of the lead levels due to the traced-out reservoirs.

Let us now look at the time dependence of equal-time expectation values $\braket{A(t)}$. Tracing out the reservoir in Eq.\,(\ref{eq: equal time full}) yields $\braket{A(t)}=\text{tr}_S\left(A\,\rho(t)\right)$, where the time-evolution of the density matrix $\rho(t)=\text{tr}_R\left(\rho_\text{full}(t)\right)$ of the system $S$ is now given by the Lindblad equation (\ref{eq: Lindblad quadratic})\,. 
Using Eq.\,(\ref{eq: Lindblad quadratic}) and the cyclicity of the trace, the time-evolution of equal-time expectation values is given by
\begin{multline}\label{eq: time evolution expect val operator}
\ii\frac{\dd}{\dd t} \braket{A(t)}=\braket{[A,\tilde{H}](t)}\\
+\ii\sum_{mn}\Lambda_{mn}^{(1)}\braket{\left(2L_m^\dagger A L_{n}^\pdag -\left\{A, L_{m}^\dagger L_{n}^\pdag \right\}\right)(t)}\\
+\ii\sum_{mn}\Lambda_{mn}^{(2)}\braket{\left(2L_n^\pdag A L_{m}^\dagger-\left\{A , L_{n}^\pdag L_{m}^\dagger\right\}\right)(t)}
\end{multline}
where each argument $t$ refers to the full operator enclosed in the foregoing brackets. 

Next we turn to correlators of the form (\ref{eq: correlators full}). Tracing out the reservoir yields $\braket{A(t)\C}=\text{tr}_S\left(A\,{\rhotilde}_{\C}(t)\right)$ with ${\rhotilde}_{\C}(t)=\text{tr}_R({\rhotilde}_{\C,\text{full}}(t))$.  Although ${\rhotilde}_{\C,\text{full}}(t)$ and $\rho_\text{full}(t)$ have the same Hamiltonian dynamics, the Liouville equation for ${\rhotilde}_{\C}(t)$ after tracing out the reservoirs differs by sign factors from that of
$\rho(t)$. This is due to the fact that the operator $\C$ in Eq.\,(\ref{eq: tilde rho def}) contains an odd number of fermionic operators, so that the standard version of the quantum regression theorem\cite{Breuer2002,Gardiner2000}, which assumes $\C$ to be bosonic, does not apply. The fermionic version of this theorem, proven in Appendix~\ref{sec: appendix B}, leads to the following time evolution for $\rhotilde_{\C}(t)$:
\begin{multline}\label{eq: rhotilde_dot}
\dot{{\rhotilde}}_{\C}(t)=-\ii[\tilde{H},{\rhotilde}_{\C}(t)]\\
+\sum_{mn}\Lambda_{mn}^{(1)}\left(\zeta\,2L_{n}^\pdag{\rhotilde}_{\C}(t)L_{m}^\dagger-\left\{L_{m}^\dagger L_{n}^\pdag,{\rhotilde}_{\C}(t)\right\}\right)\phantom{\,,}\\
+\sum_{mn}\Lambda_{mn}^{(2)}\left(\zeta\,2L_{m}^\dagger{\rhotilde}_{\C}(t)L_{n}^\pdag-\left\{L_{n}^\pdag L_{m}^\dagger,{\rhotilde}_{\C}(t)\right\}\right)\,,
\end{multline}
with $\zeta=+1 (-1)$ if $\C$ contains an even (odd) number of fermion operators. Using (\ref{eq: rhotilde_dot}) and the cyclicity of the trace, one obtains the following equation for $t>0$
\begin{multline}\label{eq: time evolution correlators}
\ii\frac{\dd}{\dd t}\braket{A(t)\C}=\braket{[A,\tilde{H}](t)\,\C}\\
+\ii\sum_{mn}\Lambda_{mn}^{(1)}\braket{\left(\zeta\,2L_{m}^\dagger A L_{n}^\pdag -\left\{A,L_{m}^\dagger L_{n}^\pdag \right\}\right)(t)\,\C}\\
+\ii\sum_{mn}\Lambda_{mn}^{(2)}\braket{\left(\zeta\,2L_{n}^\pdag AL_{m}^\dagger-\left\{A,L_{n}^\pdag  L_{m}^\dagger\right\}\right)(t)\,\C}\,.
\end{multline}

Analogously, the time dependence of $\braket{\C A(t)}$ can be obtained using $\braket{\C A(t)}=\text{tr}\left(A{\rhotilde}_{\C}'(t)\right)$, where ${\rhotilde}_{\C}'=\rhotilde {\C}$ has the same dynamics as ${\rhotilde}_{\C}$, which is given in Eq.\,(\ref{eq: rhotilde_dot}). 

Starting from Eq.\,(\ref{eq: time evolution correlators}) and the analogous equation for $\braket{\C A(t)}$ it is straightforward to set up the equations of motion for nonequilibrium Green's functions. The definitions (\ref{eq: GR definition})-(\ref{eq: G Fourier}) hold for the full system with Hamiltonian dynamics before tracing out the reservoir $R$. Therefore, they are valid for positive and negative times $t$. However, the derivation of the Lindblad equation assumes $t>0$. Thus, we will use it to evaluate $\mathcal{G}^R(t)$ and $\mathcal{G}^K(t)$ only for positive times and then use the general relations (\ref{eq: time invariance relations}) to obtain results for negative times. 

For the equation of motion of the retarded Green's function (\ref{eq: GR definition}), one obtains
\begin{align}
\ii\frac{\dd}{\dd t} \mathcal{G}_{mn}^R(t)&=\delta(t)\delta_{mn}+\sum_{k}\left(\tilde{h}_{mk}-\ii\Lambda_{mk}^{(+)}\right)\mathcal{G}^R_{kn}(t)\,,
\intertext{where we defined}
\label{eq: lambda_pm}\Lambda^{(\pm)}&=\Lambda^{(1)}\pm\Lambda^{(2)}\,.
\end{align}
Fourier transforming we obtain as final result in matrix notation:
\begin{align}
\label{eq: GR}G^R(\omega)=&\left(\omega-\tilde{h}+\ii\Lambda^{(+)}\right)^{-1}\,.
\end{align}

The equation of motion of $\mathcal{G}^K(t)$ for $\displaystyle{t>0}$ is given, via Eq.\,(\ref{eq: time evolution correlators}) and the corresponding equation for $\braket{\C A(t)}$, by
\begin{align}
\ii\frac{\dd}{\dd t}\mathcal{G}^K(t)&=\left(\tilde{h}-\ii\Lambda^{(+)}\right) \mathcal{G}^K(t)\,,\quad(t>0)\,,
\intertext{with the formal solution}
\label{eq: GK t>0} \mathcal{G}^K(t)=&\,\text{exp}\left(-\ii \tilde{h}t-\Lambda^{(+)} t\right)\mathcal{G}^K(0)\,,\quad(t>0).
\intertext{For negative times, we use Eq.\,(\ref{eq: time invariance relations}) to obtain}
\label{eq: GK t<0} \mathcal{G}^K(t)=&\,\mathcal{G}^K(0)\,\text{exp}\left(-\ii \tilde{h}t+\Lambda^{(+)}t\right)\,,\quad(t<0).
\end{align}
To find an expression for $\displaystyle{\mathcal{G}^K(0)}$, we rewrite it as
\begin{align}
\label{eq: GK expressed with P}\mathcal{G}^K(0)&=\,\ii\mathbbm{1}-2\ii P(0)\,,\quad P_{mn}(t)=\braket{L_m(t)L_n^\dagger(t)}_\text{NESS}\,.
\end{align}
Since $P_{mn}(t)$ is an equal-time expectation value, its time evolution is described by Eq.\,(\ref{eq: time evolution expect val operator}). Its time derivative is zero in the steady state because then equal-time expectation values are stationary. This implies
\begin{flalign}
 &0=\,\ii\frac{\dd}{\dd t} P(t)=\left[\tilde{h},P(t)\right]-\ii\left\{\Lambda^{(+)},P(t)\right\}+2\ii\Lambda^{(1)}\,.
\intertext{Evaluated at $t=0$, this is equivalent to}
\label{eq: equation for G(t=0)}&2\Lambda^{(-)}=\left[\mathcal{G}^K(0),\tilde{h}\right] +\ii\left\{\Lambda^{(+)},\mathcal{G}^K(0)\right\}\,.
\end{flalign}
Equation~(\ref{eq: equation for G(t=0)}) is an implicit relation for $\mathcal{G}^K(0)$. Calculating the Keldysh Green's function in Fourier space we use Eq.\,(\ref{eq: GK t>0}) for $t>0$ and Eq.\,(\ref{eq: GK t<0}) for $t<0$:
\begin{align}
 &\notag G^K(\omega)=\int_{-\infty}^{\infty}\dd t\,\e^{\ii\omega t}\mathcal{G}^K(t)\\
\notag=&\,\ii\left(\omega- \tilde{h}+\ii\Lambda^{(+)}\right)^{-1}\mathcal{G}^K(0)-\ii \mathcal{G}^K(0)\left(\omega - \tilde{h}-\ii\Lambda^{(+)}\right)^{-1}\\
\notag=&\,\ii\left(\omega- \tilde{h}+\ii\Lambda^{(+)}\right)^{-1}\left[\mathcal{G}^K(0)\left(\omega-\tilde{h}-\ii\Lambda^{(+)}\right)\right.\\
\notag&\left.-\left(\omega- \tilde{h}+\ii\Lambda^{(+)}\right) \mathcal{G}^K(0)\right]\left(\omega- \tilde{h}-\ii\Lambda^{(+)}\right)^{-1}\\
=&-\ii\left(\omega- \tilde{h}+\ii\Lambda^{(+)}\right)^{-1}2\Lambda^{(-)}\left(\omega- \tilde{h}-\ii\Lambda^{(+)}\right)^{-1}\,,
\end{align}
where we made use of Eq.\,(\ref{eq: equation for G(t=0)}) in the last step.
Comparing this with our result for the retarded Green's function (\ref{eq: GR}) we get as the final result for the Keldysh Green's function
\begin{align}
\label{eq: GK} G^K(\omega)=&-\ii G^R(\omega)\,2\Lambda^{(-)}\,G^A(\omega)\,,
\end{align}
where we exploited the Hermiticity of $\Lambda^{(+)}$.

Let us now compare the results of the Lindblad approach for $G^R(\omega)$ and $G^K(\omega)$, Eqs.\ (\ref{eq: GR}) and (\ref{eq: GK}), to those of an exact treatment of the full Hamiltonian dynamics, Eqs.\ (\ref{eq: GR Keldysh}) and (\ref{eq: GK Keldysh}). We observe that the retarded and Keldysh components of the self-energy, which in the present context of quadratic models describe the hybridization between system $S$ and reservoir $R$, are replaced by the Lindblad driving rates:
\begin{subequations}\label{eq: Sigma-lambda}
\begin{align}
&\Sigma_\text{exact}^R(\omega)\overset{\text{Lindblad}}{\to}\Delta^{\rm{Lamb}}-\ii\Lambda^{(+)}\,,\\
&\Sigma_\text{exact}^K(\omega)\overset{\text{Lindblad}}{\to}-2\ii\Lambda^{(-)}\,.
\end{align}
\end{subequations}
Of course, the matrices $\Lambda^{(\pm)}$ are independent of $\omega$ and, therefore, a finite number of Lindblad operators cannot capture the full $\omega$-dependence of a continuous self-energy $\Sigma_\text{exact}(\omega)$ in general. Nevertheless, for quantum impurity models, it will in fact be possible to capture all relevant information from the reservoirs in terms of suitably chosen Lindblad rates.

In thermal equilibrium, $\Sigma_\text{exact}^K(\omega)$ and $\Sigma_\text{exact}^R(\omega)$ are linked via the fluctuation-dissipation theorem \cite{Kamenev2011}:  
\begin{align}
\label{eq: fluctuation-dissipation}\Sigma_\text{exact}^K(\omega)&=2\ii\left(1-2f(\omega)\right)\text{Im}\left(\Sigma_\text{exact}^R(\omega)\right)\,,
\end{align}
with $f(\omega)$ being the Fermi distribution function. Hence, if the Lindblad reservoirs are used to thermalize a system, the ratio of the two matrices $\Lambda^{(\pm)}$ has to encode the details of the occupation numbers as will be elaborated below, see Eq.\,(\ref{eq: lambda(1/2)}). Let us stress, however, that due to the fact that a finite number of Lindblad operators cannot describe the full $\omega$-dependence of the self-energy, the fluctuation-dissipation theorem is, in general, not obeyed in the Lindblad approach.

Equations (\ref{eq: GR}) and (\ref{eq: GK}) are the main results of this section. They allow steady-state Green's functions for quadratic models characterized by a Lindblad equation to be calculated by simply evaluating matrix equations. These formulas have been found before \cite{PhysRevLett.110.086403,PhysRevB.89.165105} using a superoperator representation. Our derivation has the instructive feature of using only the basic definitions and relations of a Lindblad system together with the definitions of the Green's functions and their time evolution.

\section{A Lindblad approach to impurity models}\label{sec: Lindblad approach to impurity systems}
Let us now turn to impurity models. We consider models which consist of an arbitrary impurity coupled to different noninteracting fermionic leads, labeled by $\alpha$. For convenience, we will include the spin index into the channel index $\alpha$. For two spinful channels, for example, $\alpha\in\{L\uparrow,L\downarrow,R\uparrow,R\downarrow\}$, where $L$ and $R$ denote the left and right channels, respectively. Our aim is the correct description of all impurity properties in steady-state nonequilibrium that arises when different leads are held at different but fixed temperatures or chemical potentials. We consider a Lindblad approach suitable for such systems and, using the formulas for Green's functions from the previous section, we will explain in which limits our Lindblad approach reproduces the correct impurity physics. The same Lindblad equation has been suggested and used in Refs.~[\onlinecite{1.3548065,1.3581098,0953-8984-24-22-225304,1751-8121-47-9-095002,1751-8121-48-1-015004,PhysRevB.86.125111}]. We revisit it here to analyze explicitly in which limits the Lindblad equation reproduces an exact representation of a continuous reservoir, and to gain a deeper understanding of the resulting hybridization. This will be helpful in finding a local setup for MPS-based methods in Sec.~\ref{sec: Local chain}.

\subsection{Hamiltonian for impurity and leads}
The Hamiltonian of system $S$ consisting of an impurity, leads, and impurity-lead-hybridization is given by
\begin{align}
\label{eq: H_model} H =& H_\text{imp} + H_\text{lead} + H_\text{hyb}\,.
\end{align}
The impurity Hamiltonian $H_\text{imp}$ does not contain lead operators, but is otherwise arbitrary. In particular, $H_\text{imp}$ does not need to be a quadratic Hamiltonian but can contain interactions. $H_\text{lead}$ represents the noninteracting leads
\begin{align}
\label{eq: H_lead}H_\text{lead}=&\sum_{\alpha k}\varepsilon_{\alpha k}c_{\alpha k}^\dagger c_{\alpha k}=\sum_{q}\varepsilon_qc_{q}^\dagger c_{q}^\pdag\,,
\end{align}
where $q = \{\alpha,k\}$ is a composite index. If $i=1\dots M_d$ discrete impurity levels couple linearly to these fermionic leads, the general form of the hybridization between the impurity and the leads is given by
\begin{align}
\label{eq: H_int}H_\text{hyb}=&\sum_{i=1}^{M_d}\sum_{q}\left(v_{iq}\,d_{i}^\dagger c_q^\pdag + \text{h.c.}\right)\,.
\end{align} 

It is well-known that for quantum impurity models all lead properties relevant for determining the impurity self-energy are encoded in the so-called hybridization function, a matrix of dimension $M_d$ which for one lead $\alpha$ is given by
\begin{subequations}\label{eq: definition hybridization}
\begin{align}
\label{eq: definition Delta}\Delta^{R/K}_{ij,\alpha}(\omega)&=\sum_{k}v^\pstar_{iq}v^*_{jq}\,g^{R/K}_{qq}(\omega)\,.
\intertext{Here $\displaystyle{g^{R/K}_{qq}(\omega)}$ is the bare Green's function of lead level $q$ in the \textit{absence} of the coupling to the impurity. For the retarded component it suffices to consider only its imaginary part,}
\label{eq: definition Gamma}\Gamma_{ij,\alpha}(\omega)&=-\text{Im}\left(\Delta_{ij,\alpha}^{R}(\omega)\right),
\end{align}
\end{subequations}
since its real part can be deduced from the Kramers-Kronig relation. Let us also define the total hybridization
\begin{align}
\Delta^{R/K}_{ij}(\omega)&=\sum_\alpha\Delta^{R/K}_{ij,\alpha}(\omega)\,,\quad\Gamma_{ij}(\omega)=\sum_\alpha\Gamma_{ij,\alpha}(\omega)\,.
\end{align}
By definition, quantum impurity models assume continuous leads (CL), i.e.\ they assume the spectrum of lead excitations $\varepsilon_q$ to form a continuum. The bare lead correlators are assumed to describe thermal leads and hence have the well-known form
\begin{subequations}\label{eq: gRK CL}
\begin{align}
\label{eq: gR Kel}g_{qq;\text{CL}}^R(\omega)=&\left(\omega-\varepsilon_q+\ii\epsilon\right)^{-1}\,,\\
  \label{eq: gK Kel} g_{qq;\text{CL}}^K(\omega)=&-2\ii\left(1-2f_\alpha(\omega)\right)\pi\,\Lorentz{\epsilon}{q}\,.
\end{align}
\end{subequations}
Here $f_\alpha(\omega)=\left[\e^{\left(\omega-\mu_\alpha\right)/T_\alpha}+1\right]^{-1}$ is the Fermi function for decoupled lead $\alpha$ at temperature $T_\alpha$ and chemical potential $\mu_\alpha$. (When the energy argument of the Fermi function is discrete, as in $f_\alpha(\varepsilon_q)$, its index $\alpha$ will be understood to be the same as in $q= \{\alpha, k\}$.) In Eq.\,(\ref{eq: gK Kel}),  we introduced the abbreviation
\begin{align}
\label{eq: Lorentz definition}\Lorentz{\epsilon}{q}=&\frac{\epsilon/\pi}{\left(\omega-\varepsilon_{q}\right)^2+\epsilon^2}\,,
\end{align}
which we will use henceforth for a normalized Lorentz function of width $\epsilon$. When taking the continuum limit, the order of limits is such that the level spacing is sent to zero first, followed by taking $\epsilon$ to zero. Thus, in the above Eqs.\,(\ref{eq: gRK CL}) and (\ref{eq: Lorentz definition}), $\epsilon$ is an infinitesimal parameter, so that $\Lorentz{\epsilon}{q}$ becomes a true Dirac delta function.

\subsection{Lindblad equation for impurity models}\label{sec: Lindblad equation for impurity models}

\begin{figure}[tb]
\includegraphics[width=0.47\textwidth]{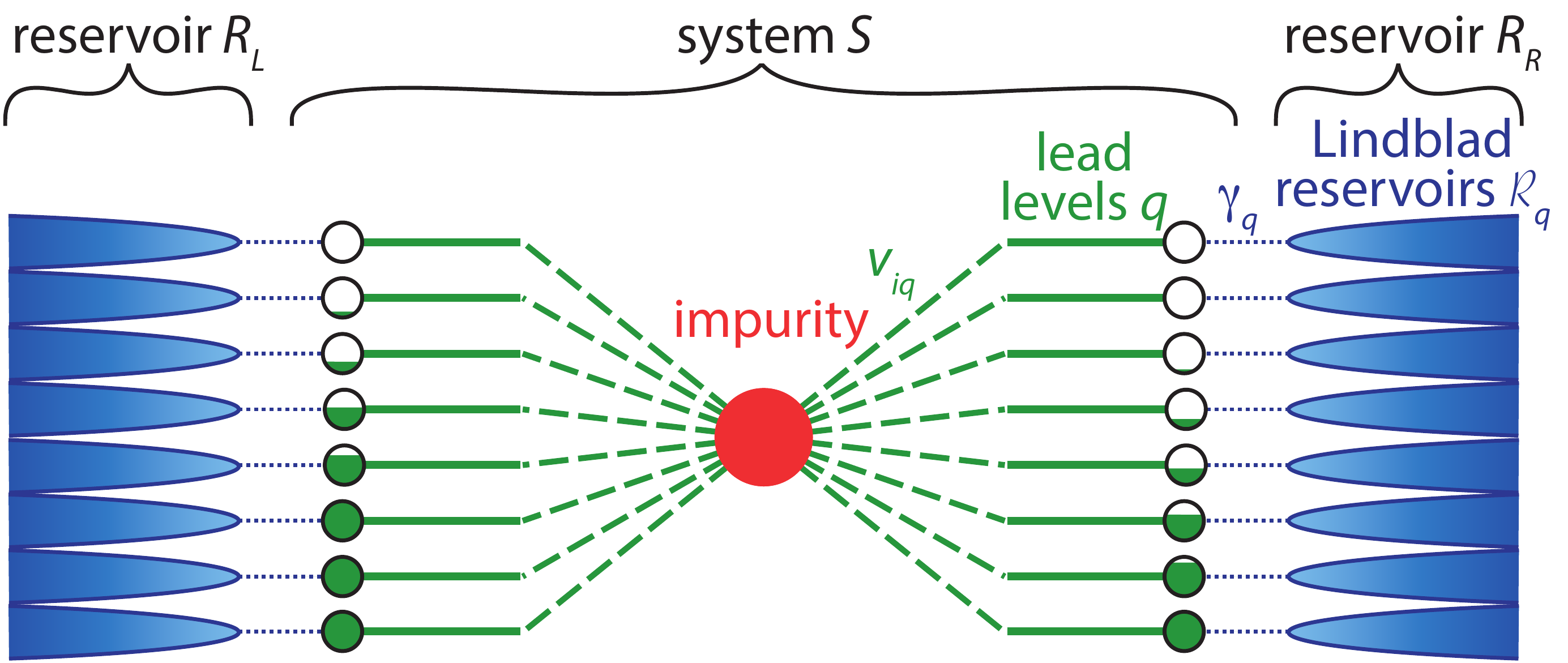}
\caption{Schematic depiction of the model for two physical leads $\alpha=\{L,R\}$. Each lead level $q$ couples to the impurity level~$i$ with coupling strength $v_{iq}$. The reservoir $R=R_L+R_R$ consists of one Lindblad reservoir $\mathcal{R}_q$ for each lead level $q$, whose Lindblad driving rate is chosen such that it tends to drive that level's occupancy towards $f_\alpha(\varepsilon_q)$ (though a small deviation from the latter will be induced by the level-dot coupling, see Sec.~\ref{sec: current RLM} for details). The value of $f_\alpha(\varepsilon_q)$ is symbolized by the degree of filling of the corresponding open circle. The occupation numbers for the left and right leads differ for a system in nonequilibrium.}
\label{fig: Sketch}
\end{figure}

The goal of the LDDL scheme is to mimic the CL description as well as possible while using a finite number of discrete lead levels. [The index $q$ is thus understood to be discrete within the context of the discrete leads (DL) in the LDDL scheme, and continuous only when referring to CL expressions.] However, a finite number of discrete lead levels is only capable of describing steady-state nonequilibrium if some dissipative dynamics is introduced that ensures that the level occupancies $N_{q;\text{DL}}(t) = \langle c^\dagger_q c^\pdag_q \rangle$ are driven towards the values $f_\alpha (\varepsilon_q)$ characteristic for the bare, uncoupled leads.  
The LDDL scheme achieves this by coupling each physical lead level $q$ to one auxiliary reservoir $\mathcal{R}_q$, as depicted in Fig.~{\ref{fig: Sketch}}, whose properties are tuned such that the dissipative dynamics of the reservoir-level system (without impurity) drives $N_{q;\text{DL}}(t)$ towards the desired value:
\begin{align}
\label{eq: limit N_q}\lim_{t\to\infty}N_{q;\text{DL}}(t)=f_\alpha(\varepsilon_q) \; . 
\end{align}
Technically, we imagine tracing out the auxiliary reservoirs and describing their effects on the discrete levels of the discretized leads using suitably chosen Lindblad terms in a Liouville equation for the system $S$ consisting of impurity plus physical leads. Note that it is not possible to use Lindblad terms to describe the dissipative effects of leads \textit{directly} coupled to the impurity, because this coupling can be strong, so that the leads cannot be treated as a Markovian bath. In contrast, as will become clear later (see Secs.~\ref{sec: hybridization} and \ref{sec: current RLM}), the couplings between the proposed Lindblad reservoirs and the lead levels go to zero in the continuum limit of infinitely many lead levels $q$. In this case, the approximations made to obtain the Lindblad equation are justified.

We now specify the Lindblad dynamics intended to ensure that the occupation of the lead levels is driven towards the steady-state values  of $N_{q;\text{DL}}(t\to\infty)=f_\alpha(\varepsilon_q)$. To this end, we first look at one lead level $q$ without coupling to the impurity $(H_q=\varepsilon_q c_q^\dagger c_q^\pdag)$ but coupled to its Lindblad reservoir $\mathcal{R}_q$. The dissipative terms in the Liouville equation are of the form
\begin{align}
\notag\mathcal{D}\rho(t)=\,&\lambda_{q}^{(1)}\left(2c_{q}^\pdag\rho(t)c_{q}^\dagger-\left\{c_{q}^\dagger c_{q}^\pdag,\rho(t)\right\}\right)\\
\label{eq: dissipation single level}+&\lambda_{q}^{(2)}\left(2c_{q}^\dagger\rho(t)c_{q}^\pdag-\left\{c^\pdag_{q}c_{q}^\dagger,\rho(t)\right\}\right)\,,
\end{align}
where $\lambda_q^{(1,2)}$ is the only entry of the matrix $\Lambda^{(1,2)}$, which in the present context is a $1\times 1$ matrix.

In this case, Eq.\,(\ref{eq: time evolution expect val operator}) (without Lambshift) can be used to determine the time evolution of the occupation number~$N_{q;\text{DL}}(t)$: 
\begin{align}
\label{eq: dgl N_q}\frac{\dd}{\dd t}N_{q;\text{DL}}(t)=2\lambda_{q}^{(2)}-2\left(\lambda_{q}^{(1)}+\lambda_{q}^{(2)}\right)N_{q;\text{DL}}(t)\,.
\end{align}
The resultant steady-state value of $N_{q;\text{DL}}(t)$ is given by
\begin{align}
\label{eq: limit N_q in terms of lambda}\lim_{t\to\infty}N_{q;\text{DL}}(t)=\frac{\lambda_{q}^{(2)}}{\lambda_{q}^{(1)}+\lambda_{q}^{(2)}}\,.
\end{align}
The requirement in Eq.\,(\ref{eq: limit N_q}), therefore, leads to
\begin{align}
\label{eq: lambda(1/2)}\lambda_{q}^{(1)}=\gamma_q\left(1-f_\alpha(\varepsilon_q)\right) \text{ and } \lambda_{q}^{(2)}=\gamma_q f_\alpha(\varepsilon_q)\,.
\end{align}
Here, $\gamma_q$ is an overall constant on the right-hand side of Eq.\,(\ref{eq: dgl N_q}), showing explicitly that $\gamma_q$ sets the time scale needed to reach the steady state. The same result has been found previously \cite{1.3548065} using a super-fermionic representation.
Equation~(\ref{eq: lambda(1/2)}) has a structure reminiscent of the fluctuation-dissipation theorem (\ref{eq: fluctuation-dissipation}), with $\Sigma^{R/K}(\omega)$ replaced by (\ref{eq: Sigma-lambda}) and $f_\alpha(\omega)$ replaced by $f_\alpha(\varepsilon_q)$. This analogy illustrates the limitation of the Lindblad approach due to the finite number of Lindblad operators: while the fluctuation-dissipation theorem contains the full Fermi function $f_\alpha(\omega)$, the Lindblad approach contains only the value at one single frequency, $f_\alpha(\varepsilon_q)$. The fluctuation-dissipation theorem is, therefore, not obeyed by the Lindblad approach in general. Note also that the observation that $\gamma_q$ sets the relevant time scale in this context is consistent with the fact that $\gamma_q$ plays the role of a decay rate in the retarded Green's function (\ref{eq: GR}).

This result for a single level serves as motivation for choosing the following Lindblad equation for the full quantum impurity system within the LDDL approach:
\begin{multline}\label{eq: Lindblad_model}
\dot{\rho}(t)=-\ii\left[H,\rho(t)\right]\\
 +\sum_q\gamma_q\left[\left(1-f_\alpha(\varepsilon_q)\right)\left(2c_{q}^\pdag\rho(t)c_{q}^\dagger-\left\{c_{q}^\dagger c_{q}^\pdag,\rho(t)\right\}\right)\right.\\
\phantom{+\sum_q\gamma_q\Big[}\left.+f_\alpha(\varepsilon_q)\left(2c_{q}^\dagger\rho(t)c_{q}^\pdag-\left\{c_{q}^\pdag c_{q}^\dagger,\rho(t)\right\}\right)\right]\,.
\end{multline}
$H$ is the Hamiltonian of system $S$, as defined in (\ref{eq:
  H_model})-(\ref{eq: H_int}), and the constants $\gamma_q$ describe
the total strength of the Lindblad driving on the levels $q$.

The parameters $\gamma_q$ in Eq.\,(\ref{eq: Lindblad_model}) are not yet fixed. In principle, they can be deduced by using the reaction-coordinate method \cite{1.3490188,1.3532408,1.4866769} to find an effective representation of the decoupled leads in terms of a discrete set of sites, each coupled to its own bath. To this end one divides the support of the hybridization function into different energy intervals, $\Gamma(\omega)=\sum_q\Gamma^{(0)}_q(\omega)$, and uses the reaction coordinate method to replace each of the baths $\Gamma^{(0)}_q(\omega)$ by a new lead level coupled to a new bath $\Gamma_q^{(1)}(\omega)$. One then traces out this new bath and finds the dissipative terms of the Lindblad equation (\ref{eq: Lindblad_model}), but with derived values of $\gamma_q$. These turn out to be proportional to the width (say $\delta_q$) of the energy interval, represented by level $q$, thus $\gamma_q\sim\delta_q$.

In this paper, we prefer to adopt a more phenomenological point of view, because for a future numerical treatment of the Lindblad setup, it will be useful to be able to treat $\gamma_q$ as a set of phenomenological parameters. (For example, in Sec.\,\ref{sec: Logarithmic Discretization}, we will discuss a logarithmic discretization scheme for which the choice $\gamma_q\sim\delta_q$ is not ideal.)  In this phenomenological view, the parameters $\gamma_q$ can be chosen in whichever way is convenient subject to only one requirement: the resulting hybridization function $\Delta_{ij,\alpha}^{R/K}$ must faithfully represent the original continuum form defined in Eq.\,(\ref{eq: definition Delta}).  Since the hybridization function (together with the impurity Hamiltonian $H_{\rm imp}$) fully determines the impurity self-energy, this requirement suffices to yield the correct impurity dynamics.   

The following subsections will be devoted to exploring how this requirement can be met. Let us here briefly preview our main conclusions.  In subsection \ref{sec: hybridization} we argue that the requirement can be fulfilled by choosing $\delta_q \lesssim \gamma_q $, while keeping $\gamma_q$ somewhat smaller than all other physical energy scales. In the subsequent subsections~\ref{sec: Greens functions RLM} and \ref{sec: current RLM} we then illustrate these statements explicitly within the context of the nonequilibrium resonant level model.  We find that considerable freedom of choice is available regarding the relation of $\delta_q$ to $\gamma_q$.  

Finally, let us note that the steady-state value of the difference between the actual and desired occupancies of lead level~$q$, say $\delta N_{q;\text{DL}} = N_{q;\text{DL}} - f_\alpha(\varepsilon_q)$, will in general not be zero, due to the coupling of that level to the impurity. However, we will show in subsection \ref{sec: current RLM} that one can achieve $\delta N_{q;\text{DL}} \ll 1$ by choosing $\delta_q \ll \gamma_q $ (for all levels). This in effect corresponds to the continuum limit of infinitely many lead levels with level spacing zero, in which case the Lindblad equation (\ref{eq: Lindblad_model}) becomes an exact representation of an arbitrary impurity coupled to continuous leads, with Fermi function occupations $f_\alpha(\omega)$.  However, we will argue that for the purposes of correctly describing the hybridization function and hence the impurity dynamics it is actually sufficient and computationally much more practical to choose $\delta_q \simeq \gamma_q $ (i.e.\ to fix their ratio to be of order unity).

\subsection{Hybridization}\label{sec: hybridization}

To demonstrate the suitability of the Lindblad equation (\ref{eq: Lindblad_model}) it suffices to look at the hybridization functions $\Delta_{ij,\alpha}^{R/K}(\omega)$, which involve only the bare lead Green's functions $g_{qq}^{R/K}(\omega)$. The lead Hamiltonian (\ref{eq: H_lead})  is quadratic and the Lindblad operators in Eq.\,(\ref{eq: Lindblad_model}) linear. Independent of whether or not the impurity contains interactions, we can therefore use the methods established in Section~\ref{sec: Greens functions} to derive an expression for the hybridization functions within the LDDL setup. We will compare these to the form obtained when using CL expressions.

The matrix equations (\ref{eq: GR}) and (\ref{eq: GK}) for the lead level $q$ decoupled from the impurity but including a Lindblad driving with diagonal matrices $\Lambda_{qq'}^{(\pm)}=\delta_{qq'}\lambda^{(\pm)}_{q}$ yield the following expressions for the discretized leads
\begin{subequations}\label{eq: g Lindblad lambda}
\begin{align}
\label{eq: gR Lindblad lambda}g_{qq;\text{DL}}^R(\omega)=&\left(\omega-\varepsilon_q+\ii\lambda_{q}^{(+)}\right)^{-1}\,,\\
\label{eq: gK Lindblad lambda}g_{qq;\text{DL}}^K(\omega)=&-2\ii\frac{\lambda^{(-)}_{q}}{\left(\omega-\varepsilon_q\right)^2+{\lambda_{q}^{(+)}}^2}\,.
\end{align}
\end{subequations}
Here we have
\begin{subequations}\label{eq: gRK Lindblad}
\begin{align}
\label{eq: lambda_pm_Fermi}\lambda^{(+)}_{q}&=\gamma_q,\quad \lambda^{(-)}_{q}=\gamma_q\left(1-2f_\alpha(\varepsilon_q)\right)
\end{align}
and therefore,
\begin{align}
\label{eq: gR Lindblad}g_{qq;\text{DL}}^R(\omega)=&\left(\omega-\varepsilon_q+\ii\gamma_q\right)^{-1}\,,\\
\label{eq: gK Lindblad}g_{qq;\text{DL}}^K(\omega)=&-2\ii\left(1-2f_\alpha(\varepsilon_q)\right)\pi\,\Lorentz{\gamma_q}{q}\,,
\end{align}
\end{subequations}
where $\Lorentz{\gamma_q}{q}$ describes a Lorentz function of width $\gamma_q$, as defined in Eq.\,(\ref{eq: Lorentz definition}).

Comparing $g^{R/K}_{qq}(\omega)$ from the Lindblad approach in Eq.\,(\ref{eq: gRK Lindblad}) to the corresponding expressions of the continuous leads in Eq.\,(\ref{eq: gRK CL}), we note that they have precisely the same structure, except that the Lindblad approach introduces an additional broadening $\gamma_q$: the infinitesimal broadening $\epsilon$ in the retarded Green's function of the continuous model, (\ref{eq: gR Kel}), is replaced by a finite broadening $\gamma_q$ in the Lindblad result (\ref{eq: gR Lindblad}). Similarly, the Keldysh component (\ref{eq: gK Lindblad}) contains a Lorentz peak of width $\gamma_q$ instead of the $\delta$-peak in the result of the continuous model, (\ref{eq: gK Kel}). Note that the fact that the Fermi functions of Eqs.\,(\ref{eq: gK Kel}) and (\ref{eq: gK Lindblad}) contain different arguments, is irrelevant because of the $\delta$-function in Eq.\,(\ref{eq: gK Kel}).

The hybridization $\Delta_{ij,\alpha;\text{DL}}^{R/K}(\omega)$ defined in (\ref{eq: definition Delta}) inherits this broadening from the free Green's functions $g^{R/K}_{qq;\text{DL}}(\omega)$. Explicitly, in the Lindblad approach, the negative imaginary part of $\Delta_{ij,\alpha;\text{DL}}^R(\omega)$ is a sum over a finite number of Lorentz peaks of width $\gamma_q$:

\begin{align}
\label{eq: definition Gamma_alpha}\Gamma_{ij,\alpha;\text{DL}}(\omega)=&\sum_{k}v_{iq}^\pstar v_{jq}^*\,\pi\,\Lorentz{\gamma_q}{q}\,.
\end{align}
In comparison, for standard continuous leads one obtains a sum over an infinite number of infinitely sharp $\delta$-peaks
\begin{align}
\label{eq: Gamma_Kel_delta}\Gamma_{ij,\alpha;\text{CL}}(\omega)=&\sum_kv_{iq}^\pstar v_{jq}^*\,\pi\,\Lorentz{\epsilon}{q}\,.
\end{align}
(We use the notation $\sum_k$ both when discussing the LDDL approach and for continuous leads, taking it to be understood that the continuum limit is implied for the latter, but not the former.)

Comparing Eqs.\,(\ref{eq: definition Gamma_alpha}) and (\ref{eq: Gamma_Kel_delta}), it becomes clear that $\Gamma_{\text{DL}}(\omega)$ will provide a faithful representation of $\Gamma_{\text{CL}}(\omega)$ if two conditions are satisfied: (i) To correctly explore the physical information encoded in $\Gamma_{\text{CL}}(\omega)$, the level spacings $\delta_q$ and driving rates $\gamma_q$ have to be so small that the characteristic spectral features of $\Gamma_{\text{CL}}(\omega)$ are well resolved. (ii) To obtain a smooth function for $\Gamma_{\text{DL}}(\omega)$, free from discretization artifacts, the discrete peak widths must be comparable to or larger than the level spacing,
\begin{align}
\label{eq: gamma sim delta} \delta_q \lesssim \gamma_q \,. 
\end{align}
Analogously, this also applies to the Keldysh component of the hybridization function, $\Delta^K_{ij,\alpha}(\omega)$.

\begin{figure}
\includegraphics[width=0.5\textwidth]{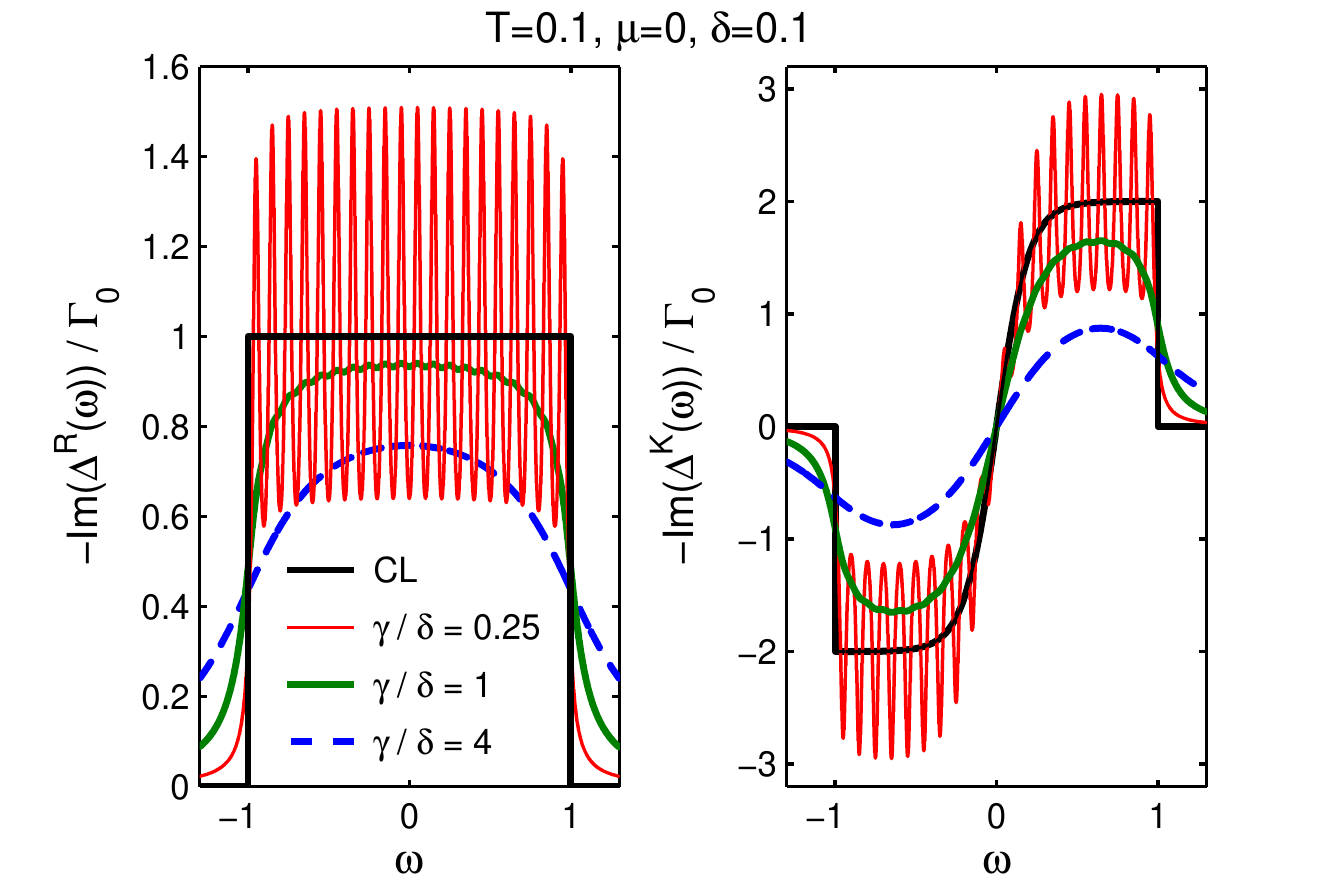}
\caption{For a single lead with a continuum hybridization function as defined in Eq.\,(\ref{eq: Gamma=box}), we plot the corresponding Lindblad result based on Eq.\,(\ref{eq: gRK Lindblad}) and the definition (\ref{eq: definition hybridization}) for different values of $\gamma$\,, which was chosen to be independent of $q$\,. A linear discretization is used with $M=2D/\delta$ lead levels for a level spacing of $\delta=0.1$\,. We need $\displaystyle{v_{k}=v=\sqrt{\Gamma_0\delta/\pi}}$ to ensure the correct continuum limit $\Delta_\text{CL}(\omega)$. The black curve represents the continuum limit (\ref{eq: Gamma=box R}) and (\ref{eq: Gamma=box K}), respectively. All energies are given in units of $D$.}
\label{fig: Hybridization}
\end{figure}

Let us illustrate this with an example. Consider a single impurity level coupled to one lead with a continuum hybridization function of the form 
\begin{subequations}\label{eq: Gamma=box}
\begin{align}
\label{eq: Gamma=box R}\Gamma_\text{CL}(\omega)=&\Gamma_0\,\theta(D-|\omega|)\,.
\end{align}
All energies are expressed in units of the half-band width~$D$. For a continuous lead in thermal equilibrium the Keldysh component $\Delta^K_\text{CL}(\omega)$ is linked to its retarded component by the fluctuation-dissipation theorem \cite{Kamenev2011}
\begin{align}
\label{eq: Gamma=box K}\Delta^K_\text{CL}(\omega)=&2\ii\left(1-2f(\omega)\right)\text{Im}\left(\Delta^R_\text{CL}(\omega)\right)\,.
\end{align}
\end{subequations}
In Fig.\ \ref{fig: Hybridization} we show the hybridization function as obtained in the Lindblad approach, which follows from inserting Eq.\,(\ref{eq: gRK Lindblad}) into Eq.\,(\ref{eq: definition hybridization}). This is done for a linear lead discretization with level spacing $\delta$ and choosing the prefactor of the Lindblad driving to be $q$-independent, $\gamma_q=\gamma$. The black curve represents the exact continuum hybridization (\ref{eq: Gamma=box}). The larger $\gamma$, the more the Lorentz peaks of Eq.\,(\ref{eq: gRK Lindblad}) are broadened. If $\gamma/\delta$ becomes too large, this leads to an unwanted smearing of the spectral features. Not illustrated in the figure, but self-evident, is the fact that this smearing can be systematically reduced by reducing the level spacing. Thus, requirement (i) can be met by choosing both $\delta$ and $\gamma$ much smaller than the relevant energy scales, here $T$, while requirement (ii) can be met by choosing $\delta \lesssim \gamma$.

Having illustrated both conditions (i) and (ii), let us remark that for \textit{equilibrium} situations, condition (ii) has a different status than condition (i). Whereas (i) is essential for getting the physics right, (ii) is needed only if one is interested in obtaining spectral properties of the impurity model, such as the local spectral function $A_d(\omega)=-\text{Im}(G_{dd}^R(\omega))$, that are more or less free from discretization artifacts. However, many physical observables, such as the linear conductance $G = (\partial J/\partial V)|_{V=0}$ through the dot or the dot occupation $N_d$, can be expressed as spectral \textit{integrals} over $A_d(\omega)$ [see Eqs.\,(\ref{eq: current_Keldysh}) and (\ref{eq: DotOccupation_Keldysh spectral function}) below]. In such cases, there is no need to avoid discretization artifacts; in fact, when using the NRG to calculate equilibrium spectral functions, it is routine practice to represent $A_d(\omega)$ as a Lehmann sum over infinitely sharp $\delta$-peaks. If necessary, it is also known empirically how to smoothen such spectral functions \cite{PhysRevLett.99.076402,RevModPhys.80.395}. To correspondingly calculate $A_{d;\text{DL}}(\omega)$ in equilibrium using the LDDL approach, it would therefore be entirely possible to choose $\gamma_q\ll\delta_q$; though this would yield a result for $A_{d,\text{DL}}(\omega)$ bearing discretization artifacts, that would not matter, because the function is integrated over anyway.

In contrast, for steady-state nonequilibrium, condition (ii) acquires additional importance, because then the Lindblad driving rates are needed to stabilize the nonequilibrium occupation functions in the leads within the transport window. Technically, they must ensure that the Keldysh component of the hybridization function (which in nonequilibrium is \textit{not} fixed by the fluctuation-dissipation theorem) is faithfully represented as a smooth function \textit{in the transport window}. To this end, it \textit{is} necessary to choose $\delta_q\lesssim \gamma_q$ within the transport window; as will be illustrated by explicit examples below, the choice $\delta_q \simeq \gamma_q$ actually suffices.

\subsection{Green's functions for the resonant level model} \label{sec: Greens functions RLM}
The hybridization function fully encapsulates all lead properties that are relevant for the impurity physics. Hence, the previous subsection constitutes a demonstration of the suitability of the suggested Lindblad equation in the context of quantum impurity models. As a check, it is instructive to explicitly calculate the impurity Green's functions for a specific quadratic model within the Lindblad approach using the methods established in Section~\ref{sec: Greens functions}. The results can be compared to the Green's functions deduced from standard Keldysh techniques using continuous thermal leads.

The simplest quadratic impurity model is the resonant level model (RLM) for spinless fermions,
\begin{align}
\begin{split}
H_\text{imp} &= \varepsilon_d c_{d}^\dagger c_{d}^\pdag\,,\\
H_\text{hyb}&=\sum_{\alpha k}v_{\alpha k}c_{d}^\dagger c_{\alpha k}^\pdag+\text{h.c.}=\sum_{q}v_{q}c_{d}^\dagger c_{q}^\pdag+\text{h.c.}\,,
\end{split}
\end{align}
where the label $d$ identifies the local level, thus $M_d=1$ in Eq.\,(\ref{eq: H_int}), and $q$ again abbreviates all lead labels, $q=\{\alpha k\}$. The RLM in the LDDL scheme as well as its continuum limit have been discussed before \cite{1.3548065} using superoperators. We revisit it here as an illustrative example of the Green's function formalism derived in Section~\ref{sec: Greens functions} and to demonstrate once more how the broadening of the Lindblad reservoirs enters the physics.

Because the RLM is quadratic, we can use equations (\ref{eq: GR}) and (\ref{eq: GK}) for the full model including the impurity and immediately write down matrix equations for the retarded Green's functions and the Keldysh Green's functions of the full system $S$. The lead-lead components of the matrices $\Lambda^{(+)}$ and $\Lambda^{(-)}$ are diagonal, $\Lambda_{qq'}=\delta_{qq'}\lambda_{q}^{(\pm)}$, with the diagonal elements  given by Eq.\,(\ref{eq: lambda_pm_Fermi}). As there is no Lindblad driving on the impurity, the matrix elements involving the local level are zero, 
\begin{align}
\Lambda^{(\pm)}_{dd}=\Lambda^{(\pm)}_{dq}=\Lambda^{(\pm)}_{qd}=0\,.
\end{align}

We first look at the retarded Green's function $\mathcal{G}^R_{mn}(t)=-\ii\theta(t)\braket{\left\{c_m^\pdag(t),c_n^\dagger\right\}}$ with $m,n\in\{d,q\}$. The matrix equation (\ref{eq: GR}) can be rewritten as
\begin{align}
\mathbbm{1}&=\left(\omega-h+\ii\Lambda^{(+)}\right)G_\text{DL}^R(\omega)\,.
\end{align}
Writing out the $dd$, $dq$, $qd$ and $qq^\prime$ components of this matrix equation separately and solving for the different correlators one readily finds
\begin{subequations}\label{eq: GR_DL RLM}
\begin{align}
\label{eq: GR_Lindblad RLM}G_{dd;\text{DL}}^{R}(\omega)=&\left(\omega-\varepsilon_d-\sum_{q}\frac{|v_{q}|^2}{\omega-\varepsilon_q+\ii\gamma_q}\right)^{-1}\,,\\
\label{eq: GR_dq} G^R_{dq;\text{DL}}(\omega)=&\left(G^A_{qd;\text{DL}}(\omega)\right)^*=\frac{v_{q}G_{dd;\text{DL}}^R(\omega)}{\omega-\varepsilon_q+\ii\gamma_q}\,,\\
\label{eq: GR_qq}G_{qq';\text{DL}}^R(\omega)=&\frac{\delta_{qq'}+v^*_{q}G_{dq';\text{DL}}^R(\omega)}{\omega-\varepsilon_q+\ii\gamma_q}\,.
\end{align}
\end{subequations}
Equation~(\ref{eq: GR_Lindblad RLM}) is consistent with (\ref{eq: gR Lindblad}), because the hybridization function $\Delta_\text{DL}^R(\omega)=\sum_q|v_q|^2 g^R_{qq;\text{DL}}(\omega)$ plays the role of the impurity self-energy here. 

Equation (\ref{eq: GK}) for the Keldysh Green's function $\mathcal{G}^K_{mn}(t)=-\ii\braket{\left[c_m^\pdag(t),c_n^\dagger\right]}$ simplifies due to the diagonal structure of $\Lambda^{(-)}$, leading to
\begin{subequations}\label{eq: GK_DL RLM}
\begin{align}
\notag &G_{dd;\text{DL}}^K(\omega)=-\ii \sum_{q}G_{dq;\text{DL}}^R(\omega)\,2\lambda_{q}^{(-)}\,G_{qd;\text{DL}}^A(\omega)\\
\label{eq: GK_Lindblad RLM}&=-2\ii\,|G^R_{dd;\text{DL}}(\omega)|^2\sum_{q} \left(1-2f_{\alpha}(\varepsilon_q)\right)|v_q|^2\,\pi\,\Lorentz{\gamma_q}{q}\,,\\
\notag&G_{qd;\text{DL}}^K(\omega)\\
\label{eq: G_kd^K}&=-\ii\sum_{q'}G_{qq';\text{DL}}^R(\omega)2\gamma_{q'}\left(1-2f_{\alpha'}(\varepsilon_{q'})\right)G_{q'd;\text{DL}}^A(\omega)\,,\\
\notag&G_{qq^{\prime};\text{DL}}^K(\omega)\\
\label{eq: G_qq^K} &=-\ii\sum_{q^{\prime\prime}}G_{qq^{\prime\prime}}^R(\omega)2\gamma_{q^{\prime\prime}}\left(1-2f_{\alpha^{\prime\prime}}\left(\varepsilon_{q^{\prime\prime}}\right)\right)G_{q^{\prime\prime}q';\text{DL}}^{A}(\omega)\,,
\end{align}
\end{subequations}
where we used Eq.\,(\ref{eq: GR_dq}).
Analogous to $G_{dd;\text{DL}}^K(\omega)$ in Eq.\,(\ref{eq: GK_Lindblad RLM}) also $G_{qd;\text{DL}}^K(\omega)$ and $G_{qq;\text{DL}}^K(\omega)$ may be expressed in terms of $G_{dd;\text{DL}}^R(\omega)$ by inserting Eqs.\,(\ref{eq: GR_DL RLM}) into Eqs.\,(\ref{eq: G_kd^K}) and (\ref{eq: G_qq^K}).

Let us now compare the $G^{R/K}_{dd}(\omega)$ correlators derived in
the Lindblad formalism to the corresponding CL expressions. The latter
are given by
\begin{align}
\label{eq: GR_Keldysh RLM}G_{dd;\text{CL}}^R(\omega)=&\left(\omega-\varepsilon_d-\sum_{q}\frac{|v_{q}|^2}{\omega-\varepsilon_q+\ii\epsilon}\right)^{-1}\,,
\end{align}
\begin{subequations}\label{eq: GK_Keldysh RLM}
\begin{align}
\notag &G_{dd;\text{CL}}^K(\omega)=\,2\ii\,\text{Im}\left(G^R_{dd;\text{CL}}(\omega)\right)\times\\
&\phantom{G_{dd;\text{CL}}^K(\omega)=\,2\ii}\sum_{\alpha}\,\left(1-2f_\alpha(\omega)\right)\frac{\Gamma_{\alpha;\text{CL}}(\omega)}{\Gamma_\text{CL}(\omega)}\\
=&\,-2\ii\,|G^R_{dd;\text{CL}}(\omega)|^2\sum_{q}\,\left(1-2f_\alpha(\varepsilon_q)\right)|v_{q}|^2\,\pi\,\Lorentz{\epsilon}{q}\,,
\end{align}
\end{subequations}
with $\Gamma_{\alpha;\text{CL}}(\omega)$ defined in (\ref{eq: Gamma_Kel_delta}) and $\Gamma_\text{CL}(\omega)=\sum_\alpha\Gamma_{\alpha;\text{CL}}(\omega)$.
Again, in Eqs.\,(\ref{eq: GR_Keldysh RLM}) and (\ref{eq: GK_Keldysh RLM}), the continuum limit is understood [as described below Eq.\,(\ref{eq: Lorentz definition})]. Comparing (\ref{eq: GR_Lindblad RLM}) with (\ref{eq: GR_Keldysh RLM}) and (\ref{eq: GK_Lindblad RLM}) with (\ref{eq: GK_Keldysh RLM}), we see explicitly that the LDDL approach reproduces the correct structure of the Green's functions, but additionally broadens the discrete lead levels to have a finite width $\gamma_q$ instead of an infinitesimal width $\epsilon$. A similar statement holds also for the $G^{R/K}_{qd}(\omega)$ and $G^{R/K}_{qq'}(\omega)$ Green's functions.

\subsection{Current and occupation functions for the resonant level model}\label{sec: current RLM}
As examples of observables for the RLM, we now calculate the current through the local level, and the occupation number of the local level and the lead levels.

\subsubsection{Current}

To determine an expression for the current through the impurity, we calculate the time derivative of the dot occupation number $N_d=\braket{c_d^\dagger c_d^\pdag}_\text{NESS}$ using Eq.\,(\ref{eq: time evolution expect val operator}). This derivative is, of course, zero, but one can identify the contributions from the different leads, $e\dot{N}_d=0=\sum_\alpha J_\alpha$. The contribution of the dissipative terms to $\dot{N}_d$ vanishes as there is no Lindblad driving on the impurity itself. Therefore, with $H_{\text{hyb};\alpha}=\sum_{k}v_{q}\,c_d^\dagger c_q^\pdag + \text{h.c.}$, we identify
\begin{flalign}
\notag J_{\alpha;\text{DL}}&=-\ii e\braket{\big[c_d^\dagger c_d^\pdag,H_{\text{hyb};\alpha}\big]}_\text{NESS}\\
\notag&=-\ii e\sum_{k}\left(v_q\braket{c_d^\dagger c_{q}^\pdag}_\text{NESS}-v_q^*\braket{c_{q}^\dagger c_d^\pdag}_\text{NESS}\right)\\
\label{eq: current, Greensfunctions}&=-e\frac{1}{4\pi}\int\dd\omega\sum_{k}\left( v_{q}G^K_{qd;\text{DL}}(\omega)+\text{h.c.}\right)\,.
\end{flalign}

Assume now that we have two leads, $\alpha=\{L,R\}$, and their hybridizations are multiples of each other, $\Gamma_\alpha(\omega)=a_\alpha\Gamma(\omega)$ with $a_L+a_R=1$ \cite{PhD_Jakobs}. We choose the discretization of both channels to be identical, $\varepsilon_{\alpha k}=\varepsilon_{k}$. This implies $|v_{\alpha k}|^2=a_\alpha|v_{k}|^2$ with $|v_{k}|^2=|v_{Lk}|^2+|v_{Rk}|^2$. In this case, it is also appropriate to set $\gamma_{\alpha k}=\gamma_{k}$. Due to $J_L+J_R=0$, we can define the current to be $J=J_L=-J_R=\left(a_RJ_L-a_LJ_R\right)$. Using Eqs. (\ref{eq: GR_dq}), (\ref{eq: GR_qq}), and (\ref{eq: G_kd^K}), one then finds for the current (with $\hbar$ restored):
\begin{multline}
\label{eq: current_Lindblad}J_\text{DL}=-\frac{4e}{h}\int\dd\omega\sum_{k}|v_{k}|^2\,\pi\,\Lorentz{\gamma_k}{k}a_La_R\\
\times\left(f_L(\varepsilon_k)-f_R(\varepsilon_k)\right)\text{Im}\left(G_{dd;\text{DL}}^R(\omega)\right)\,.
\end{multline}

The corresponding result for continuous leads is given by \cite{PhysRevLett.68.2512,PhysRevB.50.5528}
\begin{multline}
J_\text{CL}=-\frac{4e}{h}\int\dd\omega\,\Gamma_\text{CL}(\omega)a_La_R\\
\label{eq: current_Keldysh}\times\left(f_L(\omega)-f_R(\omega)\right)\text{Im}\left(G_{dd,\text{CL}}^R(\omega)\right)\,.
\end{multline}
We have seen in Section~\ref{sec: hybridization} that $\gamma_k$ should scale with the width of the energy interval $\delta_k$. Therefore, in the continuum limit of the LDDL approach, the widths of the Lorentz peaks in Eq.\,(\ref{eq: current_Lindblad}), $\gamma_k$, go to zero. In this case, we can replace $f_\alpha(\varepsilon_k)$ by $f_\alpha(\omega)$ and identify $\sum_{k}|v_{k}|^2\,\pi\,\Lorentz{\gamma_k}{k}=\Gamma_\text{DL}(\omega)$. Hence, in the continuum limit, the current in the LDDL approach has the same form as the standard CL description, while for a finite number of lead levels we recover the broadening effects discussed before.

\begin{figure}
\includegraphics[width=0.5\textwidth]{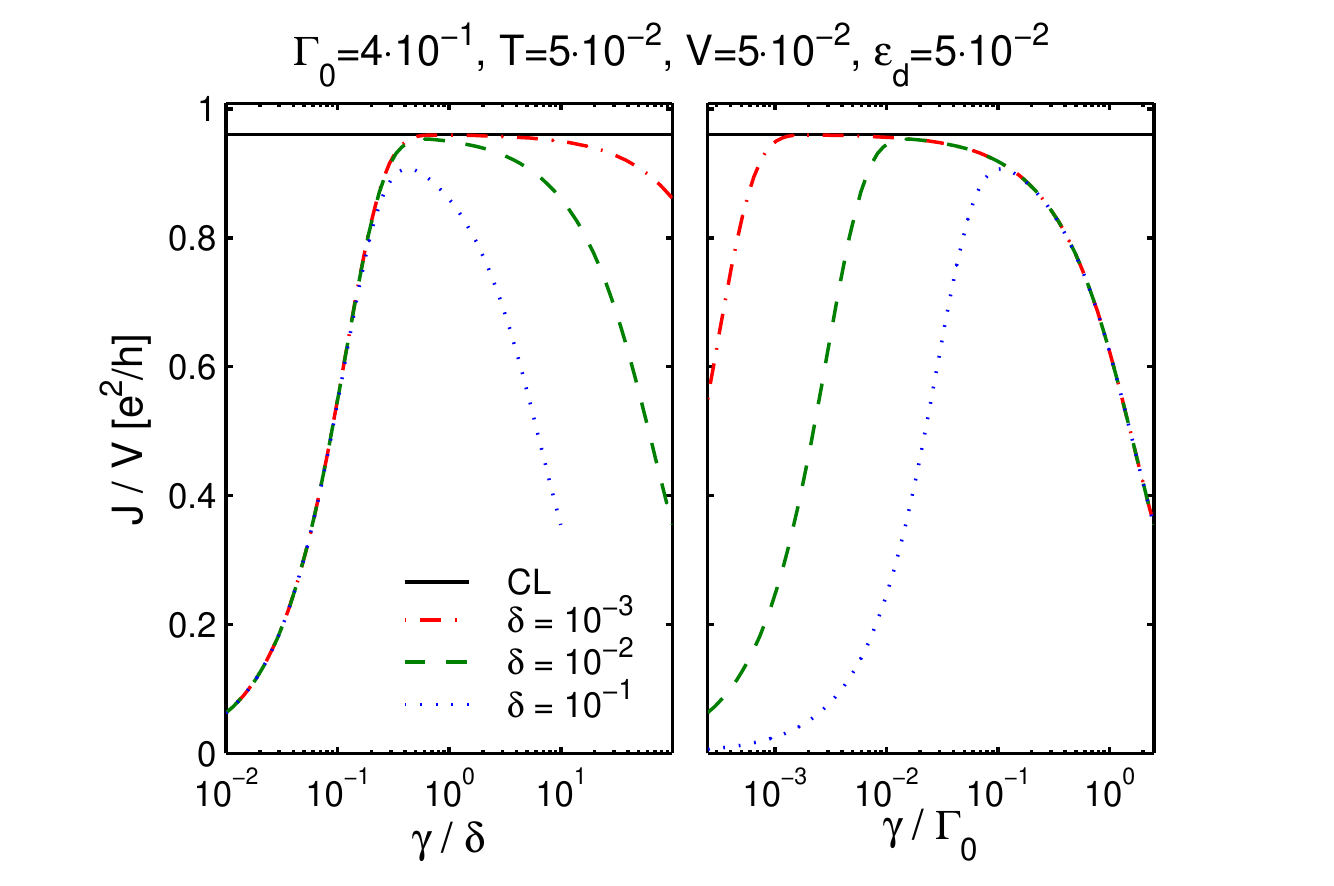}
\caption{The current through the local level of the RLM [Eq.\,(\ref{eq: current_Lindblad})] with linearly discretized leads for several values of the level spacing $\delta$. The two panels show the same data, but in the left panel as function of ${\gamma}/{\delta}$ and in the right panel as function of ${\gamma}/{\Gamma_0}$. This illustrates that the decrease in the current for small values of $\gamma$ is a discretization effect while the decrease for large $\gamma$ corresponds to an overdriving of the system. The correct physics can only be obtained if $\delta\lesssim\gamma\ll\Gamma_0$.}
\label{fig: Current}
\end{figure}

\begin{figure*}
\newlength{\mylength}
\setlength{\mylength}{0.33\linewidth}
\addtolength{\mylength}{-7mm}
\includegraphics[height=4.1cm]{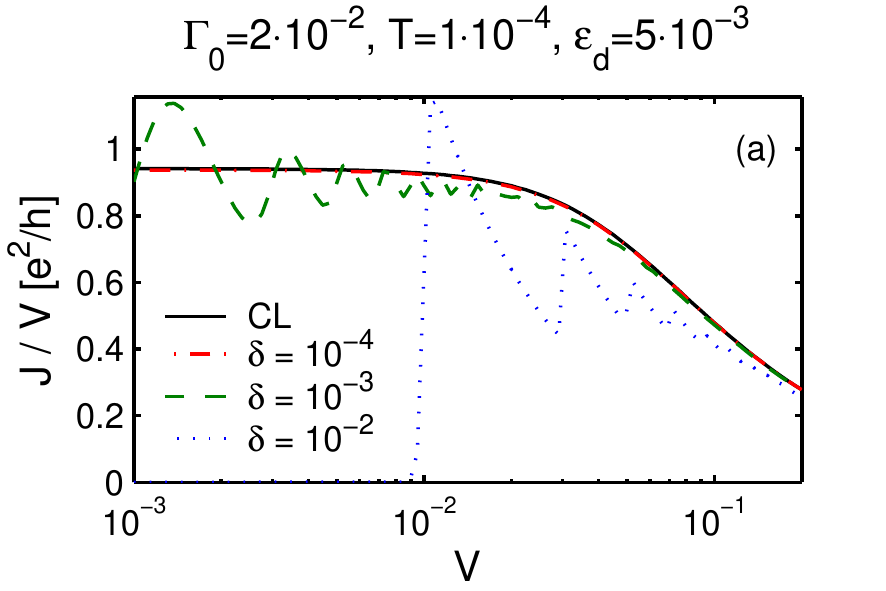}
\includegraphics[trim = 7mm 0mm 0mm 0mm, clip, height=4.1cm]{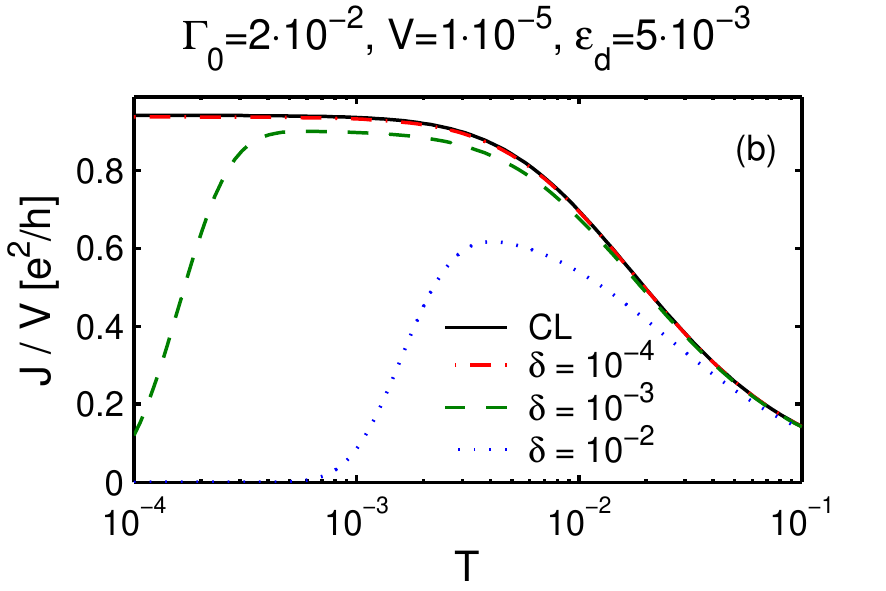}
\includegraphics[trim = 7mm 0mm 0mm 0mm, clip, height=4.1cm]{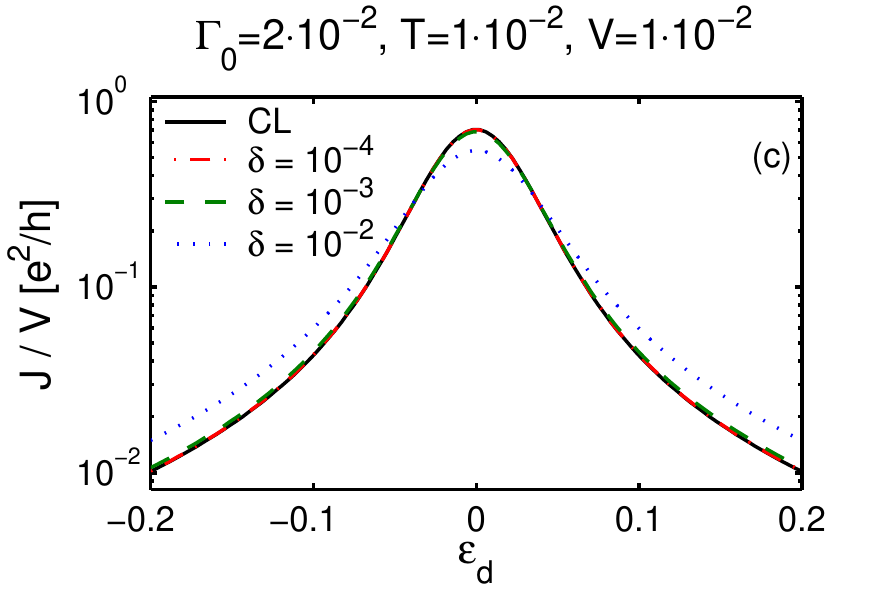}
\caption{The current [Eq.\,(\ref{eq: current_Lindblad})] for several different level spacings $\delta$ using linearly discretized leads and $\gamma_q=\delta$  as a function (a) of voltage $V$, (b) temperature $T$, and (c) the level energy $\varepsilon_d$. For sufficiently small $\delta$, the exact black curve, calculated from the continuum limit of Eq.\,(\ref{eq: current_Keldysh}), is reproduced with a deviation of less than one percent. In (a), one can clearly see discretization artifacts for $\delta=10^{-2}$ and $\delta=10^{-3}$, which vanish where $\delta/V$ gets small enough. Analogously, also in (b), it is apparent that for larger temperatures $T$, larger level spacings $\delta$ can be used, while for small $T$ small level spacings are needed.}
\label{fig: current vary parameters}
\end{figure*}

Let us illustrate the LDDL current in Eq.\,(\ref{eq: current_Lindblad}) with a few numerical examples and compare it to the exact current given by Eq.\,(\ref{eq: current_Keldysh}). We consider a symmetric continuum hybridization
\begin{subequations}\label{eq: Gamma=box two channels}
\begin{align}
\label{eq: Gamma=box two channels-couplings}
&a_L=a_R=\frac{1}{2}\,,\quad&\Gamma_\text{CL}(\omega)=\Gamma_0\,\theta(D-|\omega|)\\
\intertext{with equal temperature and symmetrically applied voltage}
&T_L=T_R=T\,,\quad&V=(\mu_L-\mu_R)=2\mu_L\,.
\end{align}
\end{subequations}
The values chosen for the different parameters can be found in the figures, where all energies are given in units of $D$\,.

In Fig.\ \ref{fig: Current} we analyze how the current through the local level, as given in Eq.\,(\ref{eq: current_Lindblad}), depends on the strength of the Lindblad driving. To this end, we discretize linearly with level spacing $\delta$ and choose $\gamma_k=\gamma$ to be \mbox{$q$-independent}. In the left panel, the current is plotted as a function of $\gamma/\delta$.  In this case, curves obtained with different level spacing $\delta$ coincide for the decrease in current when ${\gamma}/{\delta}$ decreases below $\simeq 1$, indicating that this decrease is a discretization effect. Physically, it is obvious that if $\gamma$ goes to zero, the Lindblad driving will not be able to maintain the occupation of the discretized lead levels at the values of their assigned Fermi functions. Analytically, the decrease in the current can be explained as follows: in Eq.\,(\ref{eq: current_Lindblad}) the current is expressed as an integral over the product of two peaked functions [$\text{Im}(G_{dd;\text{DL}}^R(\omega))$ and the explicit sum over $k$], whose peak positions do not precisely coincide. Therefore, if the peaks become too narrow, the integral goes to zero. To avoid this drop, one would have to broaden at least one of the two functions by hand before calculating the integral or replace the sum over $k$ by its continuum limit, $\Gamma(\omega)(f_L(\omega)-f_R(\omega))$. Such a replacement would enable one, in principle, to use arbitrarily small values of $\gamma$ in Eq.\,(\ref{eq: current_Lindblad}) for the RLM. Note, though, that it will not be possible to send $\gamma\to 0$ in Eq.\,(\ref{eq: current_Lindblad}) for more general models because a reliable calculation of the nonequilibrium Green's function $G_{dd}^R(\omega)$ will require $\gamma$ to remain finite. While for the RLM, the nonequilibrium retarded Green's function is equal to its equilibrium counterpart, this is not true in general. One will therefore need a finite broadening, $\gamma \simeq \delta$, to keep the occupation numbers of the discrete lead levels \textit{close to} the corresponding Fermi distribution (see also Sec.~\ref{sec: current RLM}) while solving for the steady state of the Lindblad equation and thereby determining the true nonequilibrium Green's function.

The second panel shows the same data as a function of $\gamma/\Gamma_0$. Here, the different curves coincide for the decrease in the current when $\gamma/\Gamma_0$ increases past $\simeq1$, illustrating that this effect is an inherent property of the Lindblad equation. It corresponds to an overdriving of the system, i.e.\ the Lindblad reservoirs destroy the coherence and hence suppress the current when $\gamma\gtrsim\Gamma_0$\,.

If the ratio $\delta/\Gamma_0$ is small enough, a plateau for $\delta\lesssim\gamma\ll\Gamma_0$ appears and the height of this plateau agrees well with the exact current obtained from Keldysh calculations.

In total, the Lindblad driving rates $\gamma_q$ must be small compared to the physical energy scale $\Gamma_0$ but larger or comparable to the level spacing $\delta_q$. On the other hand, the level spacing $\delta_q$ has to resolve the energy scale $\Gamma_0$, $\delta\lesssim\Gamma_0$. Therefore, $\gamma_q=\delta_q$ should always be an appropriate choice and we will use this choice in the following examples. 

In Figs.\ \ref{fig: current vary parameters}(a)-(c) the value of $\gamma$ is fixed to $\gamma=\delta$ and the current is plotted as a function of voltage $V$, temperature $T$, and level position $\varepsilon_d$, respectively. For small enough level spacing, the deviation from the standard continuum result represented by the black line is less than one percent. 

\begin{figure}
\includegraphics[width=0.5\textwidth]{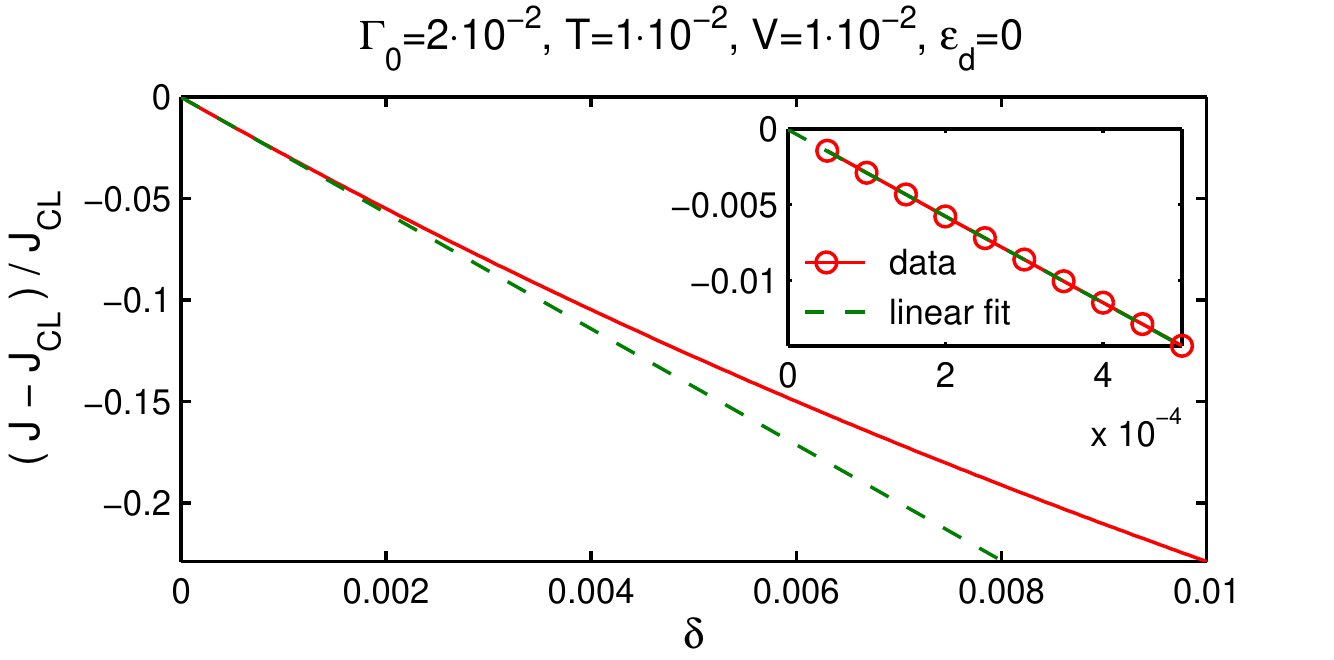}
\caption{Relative error of the current for the parameters of Fig. \ref{fig: current vary parameters}(c), at $\varepsilon_d=0$. A linear fit, obtained from the data shown in the inset, yields an offset smaller than $10^{-4}$, showing that the LDDL scheme becomes exact in the continuum limit.}
\label{fig: rel err}
\end{figure}

To be more specific, in Fig.\ \ref{fig: rel err}, we show how the relative error of the current scales with level spacing $\delta$, using $\gamma_k=\delta$. Extrapolating the data points for small $\delta$ towards the continuum limit $\delta\to 0$ using a linear fit yields an offset of the order of $10^{-4}$, demonstrating that the suggested Lindblad approach becomes exact in the continuum limit. 

In order to properly reproduce the dependence of the current on $V$, $T$, and $\Gamma_0$, the choice of level spacing must satisfy certain conditions. These can be deduced by inspecting Eq.\,(\ref{eq: current_Lindblad}), which contains an integral over the product of $\text{Im}(G_{dd;\text{DL}}^R(\omega))$ and $\sum_{k}|v_{k}|^2\pi\Lorentz{\gamma_k}{k}a_La_R\left(f_L(\varepsilon_k)-f_R(\varepsilon_k)\right)$. For $\gamma=\delta$, both these functions are smooth. Evidently, $\delta$ must be small enough to resolve the $\omega$ dependence of $G_{dd}(\omega)$ and $\Gamma(\omega)$. For the RLM, this implies that $\delta\ll\Gamma_0$ is needed. The energy scale on which $\left(f_L(\varepsilon_k)-f_R(\varepsilon_k)\right)$ varies, is set by temperature and voltage. First, consider the case that temperature is the smallest physical energy scale, $T\ll V,\Gamma_0$. $T$ sets the width of the Fermi function steps. Hence, one might expect that $\delta\lesssim T$ is needed. However, $\delta\lesssim V$ suffices. The reason is that the Fermi functions are multiplied by a smooth function, $\text{Im}(G_{dd}^R(\omega))$, which varies on an energy scale $\Gamma_0\gg T$; when integrated over, the result is independent of $T$. Note that for $V\lesssim T$ this temperature independence is lost because then the two steps of $f_L(\omega)$ and $f_R(\omega)$ are not well separated. Next consider the case  $V\ll T,\Gamma_0$. Then $\left(f_L(\omega)-f_R(\omega)\right)$ varies on an energy scale given by temperature $T$, and the voltage does not need to be resolved. Hence, in summary, $\delta$ has to be chosen small enough to resolve all features of the spectral function $\text{Im}(G^R_{dd}(\omega))$ and the larger of the two energy scales $V$ and $T$.

\begin{figure}[b]
\includegraphics[width=0.5\textwidth]{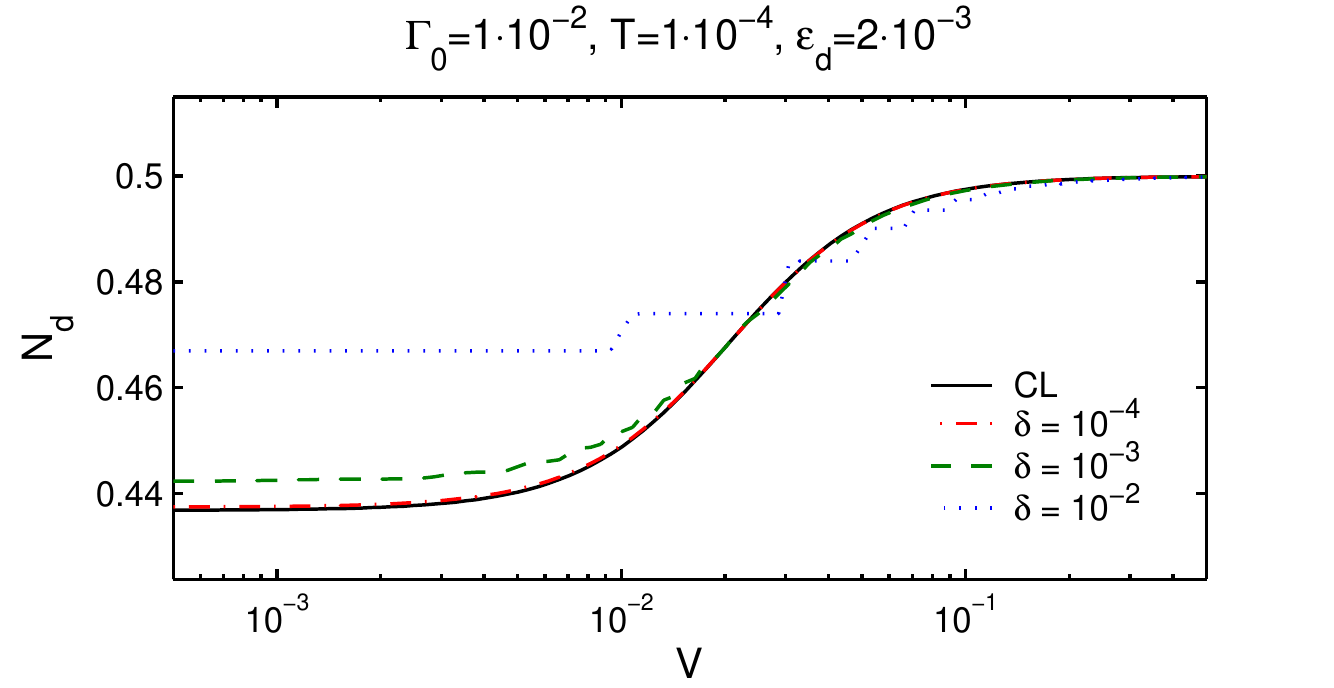}
\caption{The occupation number of the local level for the RLM in the Lindblad approach as given by Eq.\,(\ref{eq: DotOccupation_Lindblad}), as function of a symmetrically applied voltage. The discretization was chosen to be linear with different values for the level spacing $\delta$\,. The black curve represents the continuum limit of Eq.\,(\ref{eq: DotOccupation_Keldysh}). If the level spacing $\delta$ is small enough compared to the voltage $V$, the exact result is recovered. For $\delta=10^{-2}$, one can clearly see discretization artifacts.}
\label{fig: DotOccupation}
\end{figure}

\subsubsection{Occupation of local level}

The current is an observable that illustrates the dynamics of the system. As an example of a static property, we next consider the occupation number of the local level, $N_d=\braket{c_d^\dagger c_d^\pdag}_\text{NESS}$. Using the Green's functions (\ref{eq: GR_Lindblad RLM}) and (\ref{eq: GK_Lindblad RLM}) it is given by
\begin{align}
\notag &N_{d;\text{DL}}=\frac{1}{2}+\frac{1}{2\ii}\mathcal{G}^K_{dd;\text{DL}}(0)=\frac{1}{2}+\frac{1}{4\pi\ii}\int\dd\omega G^K_{dd;\text{DL}}(\omega)\\
\label{eq: DotOccupation_Lindblad}=&\frac{1}{\pi}\int\dd\omega|G_{dd;\text{DL}}^R(\omega)|^2\sum_qf_\alpha(\varepsilon_q)|v_q|^2\,\pi\,\Lorentz{\gamma_q}{q}\,,
\end{align}
where we exploited the sum rule
\begin{align}
\notag-&\frac{1}{2\pi}\int\dd\omega|G_{dd;\text{DL}}^R(\omega)|^2\sum_q|v_q|^2\,\pi\,\Lorentz{\gamma_q}{q}\\
=&\frac{1}{2\pi}\int\dd\omega\,\text{Im}\left(G_{dd;\text{DL}}^R(\omega)\right)=-\frac{1}{2}.
\end{align}
The corresponding result for continuous thermal leads is given by
\begin{subequations}\label{eq: DotOccupation_Keldysh}
\begin{align}
\label{eq: DotOccupation_Keldysh spectral function}N_{d;\text{CL}}=&-\frac{1}{\pi}\int\dd\omega\,\text{Im}\left(G^R_{dd;\text{CL}}(\omega)\right)\sum_\alpha f_\alpha(\omega)\frac{\Gamma_{\alpha;\text{CL}}(\omega)}{\Gamma_{\text{CL}}(\omega)}\\
\label{eq: DotOccupation_Keldysh peaks}=&\frac{1}{\pi}\int\dd\omega|G^R_{dd;\text{CL}}(\omega)|^2\sum_\alpha f_\alpha(\varepsilon_q)|v_q|^2\,\pi\,\Lorentz{\epsilon}{q}\,.
\end{align}
\end{subequations}
Analogously to the discussion of the current, the comparison of the Lindblad result (\ref{eq: DotOccupation_Lindblad}) to (\ref{eq: DotOccupation_Keldysh peaks}) reveals that the LDDL approach in the continuum limit recovers the standard result obtained using continuous thermal leads. 

For a symmetric hybridization of the form (\ref{eq: Gamma=box two channels}), we illustrate these formulas in Fig.\ \ref{fig: DotOccupation} where we plot the occupation of the local level given in (\ref{eq: DotOccupation_Lindblad}) as function of voltage. The discretization is again chosen linear for both leads and the Lindblad driving is set to the constant value $\gamma_q=\gamma=\delta$. Again, we find excellent agreement with the continuum results if the level spacing is chosen small enough.

\subsubsection{Occupation of lead level}

Finally, we discuss the steady-state occupation $N_{q;\text{DL}}$ of lead level~$q$.  Although our choice for the Lindblad driving rates [\Eq{eq: lambda(1/2)}] is designed to drive $N_{q;\text{DL}}$ towards its Fermi distribution value, $N_{q;\text{DL}}$ actually differs slightly from $f_\alpha(\varepsilon_q)$, due to the coupling of level $q$ to the impurity.  Using Eqs.\,(\ref{eq: GR_DL RLM}) and (\ref{eq: GK_DL RLM}), the difference can be calculated analogously to Eq.\ (\ref{eq: DotOccupation_Lindblad}), with the result:
\begin{align}
\notag &\delta N_{q;\text{DL}}=N_{q;\text{DL}}-f_\alpha(\varepsilon_q)=\\
\notag-&\int\frac{\dd\omega}{\pi}\,\frac{|v_q|^2\gamma_q\left(1-2f_\alpha(\varepsilon_q)\right)}{\left(\omega-\varepsilon_q\right)^2+\gamma_q^2}\text{Re}\left(\frac{G_{dd;\text{DL}}^R(\omega)}{\omega-\varepsilon_q+\ii\gamma_q}\right)\\
\label{eq: delta N_q}-&\int\frac{\dd\omega}{2\pi}\,\sum_{q'}\frac{\gamma_{q'}\left(1-2f_{\alpha'}(\varepsilon_{q'})\right)|v_q|^2|v_{q'}|^2\,|G_{dd;\text{DL}}^R(\omega)|^2}{\left(\left(\omega-\varepsilon_q\right)^2+\gamma_q^2\right)\left(\left(\omega-\varepsilon_{q'}\right)^2+\gamma_{q'}^2\right)}\,.
\end{align}
For the symmetric two-channel RLM as defined in Eq.\,(\ref{eq: Gamma=box two channels}), assuming that the parameters $\delta_{q'}$ and $\gamma_{q'}$ (for all $q'$) are much smaller than all other energy scales, and $\delta_{q'} \lesssim \gamma_{q'}$, this reduces to
\begin{align}
\label{eq: delta N_q RLM}\delta N_{q;\text{DL}}\simeq-\frac{|v_q|^2}{2\gamma_q}\frac{\Gamma_0\left(f_\alpha(\varepsilon_q)-f_{\bar{\alpha}}(\varepsilon_q)\right)}{(\varepsilon_q-\varepsilon_d)^2+\Gamma_0^2} \, , 
\end{align}
with $\bar{\alpha}=R(L)$ if $\alpha=L(R)$. In this case, therefore, the deviation is non-zero in the transport window where $f_\alpha \neq f_{\bar \alpha}$, and vanishes completely only for a system in equilibrium.

\begin{figure}
\includegraphics[trim = 0mm 12mm 0mm 0mm, clip, width=0.5\textwidth]{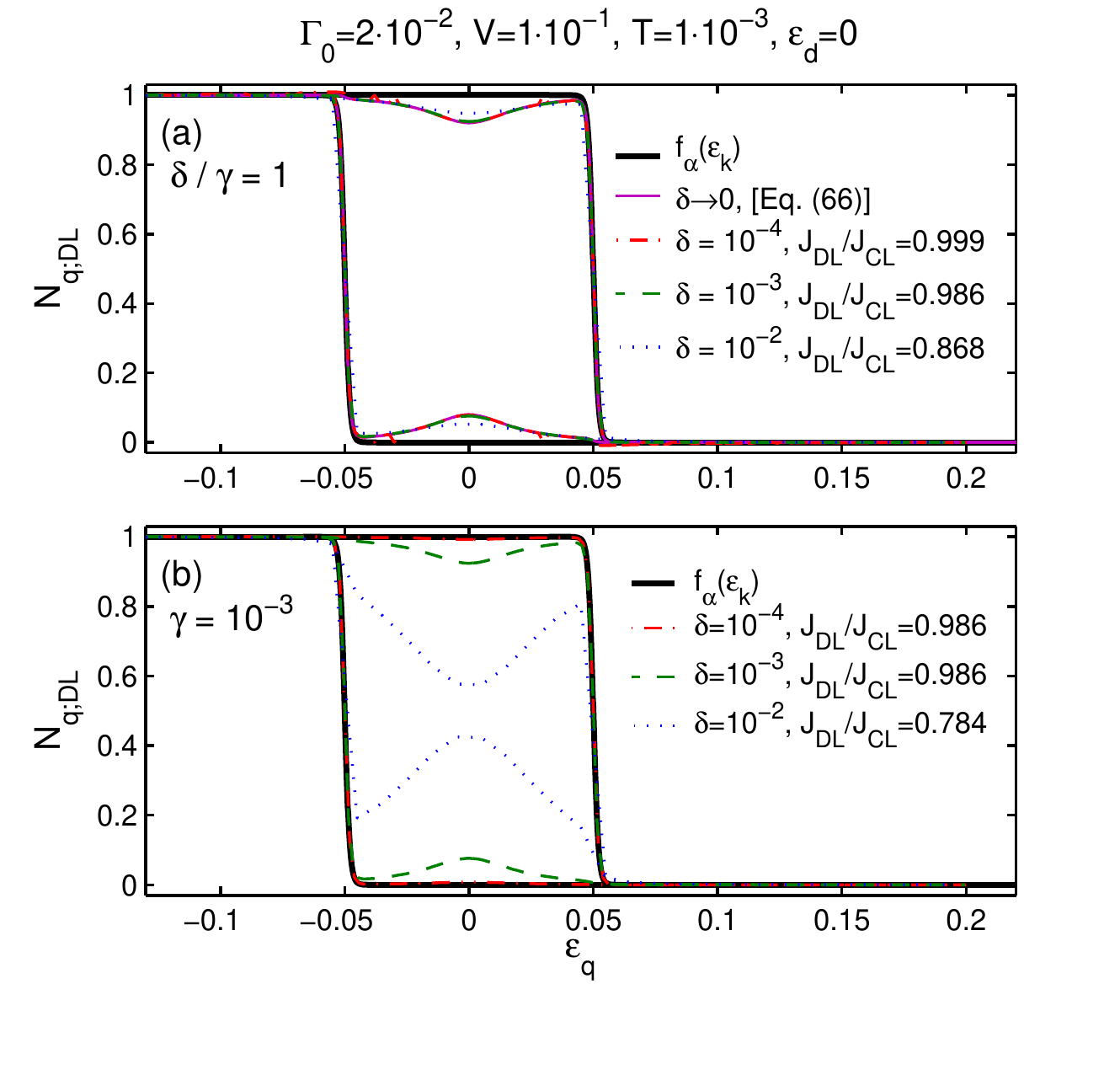}
\caption{The occupation numbers $N_{q;\text{DL}}$ for the left and right channels of a symmetric RLM as defined in Eq.\,(\ref{eq: Gamma=box two channels}), choosing a linear discretization with level spacing $\delta_q=\delta$ and $q$-independent broadening $\gamma_q=\gamma$. In (a), we show $N_{q;\text{DL}}$ for several values of $\delta$ at a fixed ratio $\delta / \gamma=1$. If the level spacing is small enough, the exact current is reproduced, although $N_{q;\text{DL}}$ deviates from the Fermi distribution by a non-zero amount $\delta N_{q;\text{DL}}$, which for small enough $\delta$ is given by Eq.\,(\ref{eq: delta N_q RLM}). In (b), $\gamma$ is kept fixed at a value that can resolve the physical relevant energy scales (here $\Gamma_0$ and $V$) while $\delta$ is varied. Reducing $\delta$, $N_{q;\text{DL}}$ approaches the Fermi distribution, but as soon as $\delta$ becomes $\lesssim\gamma$ the accuracy of $J_\text{DL}$  (taking $J_\text{CL}$ as reference) does not improve.}
\label{fig: NqVsFermi}
\end{figure}

Equation (\ref{eq: delta N_q}) is also true for more general impurity models (with $G_{dd;\text{DL}}(\omega)$ depending on the precise form of $H_{\text{imp}}$). It can be shown that the scaling of $\delta N_{q,\text{DL}}$ with $|v_q|^2/\gamma_q$ found in Eq.\,(\ref{eq: delta N_q RLM}) holds independent of the form of the impurity, again assuming $\delta_{q'}$ and $\gamma_{q'}$ small enough and $\delta_{q'}\lesssim\gamma_q$. For typical impurity models, $|v_q|^2$ is a smooth function of $q$ whose magnitude scales with the size of the corresponding energy interval, $|v_q|^2\sim\delta_q$. Therefore, if one sends both $\delta_q$ and $\gamma_q$ to zero while keeping $\delta_q \simeq \gamma_q$ (i.e.\ fixing their ratio to be of order unity), then $\delta N_{q;\text{DL}}$ does not vanish. This is depicted in panel (a) of Fig.\,\ref{fig: NqVsFermi}.

If one insists on having \mbox{$\delta N_{q;\text{DL}} \ll 1$}, one may achieve this by choosing $\delta_q\ll\gamma_q$ (thus ensuring $|v_q|^2/\gamma_q \ll 1$) while keeping $\gamma_q$ somewhat smaller than all other energy scales.  In fact, this corresponds to the order of limits used to recover the case of continuous thermal leads: first the level spacing is sent to zero and the number of lead levels to infinity while keeping the level broadening fixed and nonzero; and only subsequently the level broadening is taken to be infinitesimally small -- its only trace in the description of continuous leads is the infinitesimal damping factor $i \epsilon$ in energy denominators, e.g.\ in \Eq{eq: GR_Keldysh RLM}. Thus, for continuous leads one indeed has \mbox{$\delta N_{q;\text{CL}}=0$}, as depicted in panel (b) of Fig.\,\ref{fig: NqVsFermi}. The physical reason for this is that if the leads form a true continuum, i.e., the width of each lead level is larger than the level spacing, the effect of a single dot level on the occupation of each individual lead level is negligibly small. 

Note, however, that for numerical computations it would be impractical to use $\delta_q \ll \gamma_q$, since this would require using many more lead levels than for the case $\delta_q\simeq\gamma_q$. Moreover, when one's interest is focused only on impurity properties, it is actually \textit{not necessary} to achieve $\delta N_{q;\text{DL}} \ll 1$: in that case, the precise value of $\delta N_{q;\text{DL}}$ is irrelevant, as long as the hybridization function is represented faithfully and is smooth within the transport window. Indeed, we have shown in Sec.\,III\,C that this \textit{can} be achieved when using $\delta_q\simeq\gamma_q$, by simply taking both to be somewhat smaller than all other physically relevant energy scales.

\section{Local chain representation of the Lindblad equation}\label{sec: Local chain}

The resonant level model is a quadratic model that can be solved analytically. If the impurity contains interactions and many-particle physics becomes relevant, one can still use the suggested LDDL approach as it reproduces the correct bare hybridization function. However, in general, the Lindblad equation cannot be solved for its steady state analytically.

A versatile tool for numerical representations of many-particle quantum states are the so-called matrix product states (MPS) and matrix product operators (MPO) \cite{Schollwoeck201196}. Only recently the idea to solve Lindblad equations numerically based on MPS/MPO has gained attention: One possibility is the explicit time-evolution of the full density matrix \cite{PhysRevB.92.125145, PhysRevLett.116.237201}. Alternatively, one can step down from the level of density matrices to the level of quantum states at the price of stochastic averaging as in the stochastic quantum trajectory approach \cite{RevModPhys.70.101,Gardiner2000,Daley2014,PhysRevLett.102.040402,PhysRevLett.103.240401,PhysRevA.84.041606}. Which of the two methods is numerically less expensive strongly depends on the model and its specific parameters \cite{Bonnes2014}. To avoid the explicit time-evolution one can also target the steady state directly by solving $\dot{\rho}(t)=\mathcal{L}\rho(t)=0$ \cite{PhysRevLett.114.220601,PhysRevA.92.022116}.

MPS/MPO methods presuppose models having the structure of one-dimensional quantum chains.  If we would write our proposed Lindblad setup as a chain by simply representing each level $q=\{\alpha k\}$ by one chain site, this would result in a highly non-local model, in which each and every chain site couples to the impurity. This non-locality would render standard MPS/MPO techniques, e.g.\ for the time-evolution of a state or operator, numerically costly\footnote{In specific contexts, the added costs of this non-locality may be offset by lower entanglement, see Ref.~[\onlinecite{PhysRevB.90.235131}]}. In this section, our goal is therefore to reformulate our Lindblad scheme in such a way that the Hamiltonian and the Lindblad driving terms are local when the leads are represented by chains of the type needed for MPS/MPO calculations, where `local' means that the matrices $h$ and $\Lambda^{(1,2)}$ only connect sites on the chains that are very close to each other or are diagonal all-together.

For equilibrium calculations, it is well-known from NRG how to map the Hamiltonian of a non-interacting discretized lead onto a chain in such a way that the resulting Hamiltonian is local \cite{RevModPhys.47.773,RevModPhys.80.395} using a unitary transformation of the form $c_q=\sum_l U_{ql} c'_l$.  For our nonequilibrium LDDL scheme, however, a problem arises: under such a transformation the Lindblad matrices $\Lambda^{(1,2)}$ which in our original formulation are local ($\Lambda_{qq'}^{(1,2)}=\delta_{qq'}\lambda_q^{(1,2)}$, i.e.\ involving no driving terms that combine $c_q$ and $c_{q'}^\dagger$ for $q\neq q'$), would become strongly non-local.  The reason is that the transformed Lindblad matrices, 
\begin{align} \label{eq:map-to-chain}
  \Lambda_{ll'}^{\prime(1,2)} =\sum_qU_{ql}^* \Lambda_{qq'}^{(1,2)}
  U_{q'l'} \, , \end{align} 
would not be diagonal, because the old Lindblad matrices $\Lambda^{(1,2)}$, though diagonal, depend on $q$, e.g., due to the dependence of the diagonal elements $\lambda_q^{(1,2)}$ on the Fermi function $f_\alpha (\varepsilon_q)$.
 
This problem can be circumvented if the original Lindblad rates $\gamma_q$ are $q$-independent. To this end, we will formulate an equivalent new Lindblad equation that reproduces the same hybridization function as the one suggested in Sec.\,\ref{sec: Lindblad approach to impurity systems}, but is based on new Lindblad matrices $\tilde \Lambda^{(1,2)}$ that are proportional to the identity matrix in their $q$ indices. They are thus not only local but also invariant under arbitrary unitary transformations acting on the index $q$. This invariance makes it possible to map the leads onto a chain on which the Hamiltonian is local, without losing the locality of the dissipative Lindblad terms.  We will thus refer to the new scheme as `local setup', and to the original one as `non-local setup'.  The cost for achieving locality is that each physical lead is replaced by two auxiliary leads.  However, depending on the precise form of the impurity model, some linear combinations of auxiliary lead modes may decouple, thus lowering the cost again.

Before presenting the technical details of the local setup, let us describe its main idea. The Lindblad setup we are aiming for must have Lindblad matrices $\tilde{\Lambda}^{(1,2)}$ that are proportional to the identity matrix in their $q$ indices. They thus cannot contain any information about Fermi functions. Moreover, the occupation number towards which such matrices drive any level $q$ is actually independent of $q$ [see Eq.\,(\ref{eq: limit N_q in terms of lambda})]. The levels in the local scheme thus cannot correspond to physical levels; instead, they have the status of auxiliary levels, and Fermi-function information \textit{will have to be encoded in their coupling strengths to the impurity}. To see heuristically how such a Lindblad driving can still be used to mimic thermal leads, we note that a physical level with occupancy $f_\alpha(\varepsilon_q)$ is \textit{empty} with probability $1-f_\alpha(\varepsilon_q)$ and \textit{filled} with probability $f_\alpha(\varepsilon_q)$.  Now, occupancies of empty or filled \textit{are} describable using $q$-independent diagonal Lindblad matrices, at the cost of introducing a new index, $\eta = 1$ or 2, to distinguish the two cases.  (The matrices $\tilde{\Lambda}^{(1,2)}$ are then proportional to the identity in their $q$ indices for each $\eta$ independently. When mapping the system onto a chain the unitary transformation therefore  must not mix different $\eta$, but treat $\eta=1$ and $\eta=2$ as two independent channels.)
In the local setup we thus `double' all levels: each physical level $q$ from the non-local setup, having energy $\varepsilon_q$ and impurity coupling strength $|v_{iq}|^2$, is replaced by a pair of two auxiliary levels, $q\to\{q\eta\}$ with $\eta\in\{1,2\}$, both with the same energy $\varepsilon_q$.  We take the auxiliary level with $\eta=1$ to have coupling strength $|v_{iq}|^2[1- f_\alpha(\varepsilon_q)]$ while being Lindblad-driven towards occupancy \textit{zero}, and the auxiliary level with $\eta=2$ to have coupling strength $|v_{iq}|^2f_\alpha(\varepsilon_q)$ while being Lindblad-driven towards occupancy \textit{one}. This level-doubling construction is depicted schematically in Figs.~\ref{fig: Lead Levels}(a) and (b). As will be shown below, the local setup leads to the same hybridization function as the non-local one, and hence describes the same impurity physics.

  The Hamiltonian and Lindblad equation of the local setup have the same structure as for the non-local one [cf.\ Eqs.~\eqref{eq: H_model} to \eqref{eq: H_int} and \eqref{eq: Lindblad_model}], but with $q$ replaced by $\{q\eta\}$ and making new choices for the couplings and Lindblad driving rates. Explicitly, the Hamiltonian and impurity-lead couplings now take the form 
\begin{align}
H =H_\text{dot}+\sum_{q\eta}\left[\sum_{i=1}^{M_d}\left(\tilde{v}_{iq\eta}d_{i}^\dagger \tilde{c}^\pdag_{q \eta}+\text{h.\,c.}\right) +\varepsilon_q\tilde{c}_{q\eta}^\dagger \tilde{c}^\pdag_{q\eta}\right] \, , 
\\
\label{eq: Vk_tilde}
\tilde{v}_{iq,\eta=1} =v_{iq}\sqrt{\left(1-f_\alpha\left(\varepsilon_q\right)\right)}\,, \quad\tilde{v}_{iq,\eta=2}=v_{iq}\sqrt{f_\alpha\left(\varepsilon_q\right)}\, ,
\end{align}
As before, the Lindblad matrices are chosen diagonal, with $\tilde{\Lambda}_{q\eta,q'\eta'}^{(1,2)}=\delta_{qq'}\delta_{\eta\eta'}\tilde{\lambda}_{q\eta}^{(1,2)}$ and the Lindblad equation reads
\begin{multline}
\dot{\rho}(t)=-\ii\left[\tilde{H},\rho(t)\right]\\
 +\sum_{q\eta}\left[\tilde{\lambda}^{(1)}_{q\eta}\left(2\tilde{c}_{q\eta}^\pdag\rho(t)\tilde{c}_{q\eta}^\dagger-\left\{\tilde{c}_{q\eta}^\dagger \tilde{c}_{q\eta}^\pdag,\rho(t)\right\}\right)\right.\\
\phantom{+\sum_q\gamma_q\Big[}\left.+\tilde{\lambda}^{(2)}_{q\eta}\left(2\tilde{c}_{q\eta}^\dagger\rho(t)\tilde{c}_{q\eta}^\pdag-\left\{\tilde{c}_{q\eta}^\pdag \tilde{c}_{q\eta}^\dagger,\rho(t)\right\}\right)\right]\,.
\end{multline}
Since we want to drive the auxiliary levels with $\eta=1~(\eta=2)$ towards occupancy zero (one), they should be Lindblad-driven only by annihilation (creation) operators, respectively. Using the same Lindblad rates $\gamma_q$ for both, we thus choose
\begin{align}
\label{eq: lambda_tilde}\tilde{\lambda}_{q\eta}^{(1)}&=\delta_{\eta,1}\,\gamma_q, \quad\tilde{\lambda}_{q\eta}^{(2)}=\delta_{\eta,2}\,\gamma_q.
\end{align}
The rates $\tilde{\lambda}^{(\pm)}_{q\eta}=\tilde{\lambda}^{(1)}_{q\eta}\pm\tilde{\lambda}^{(2)}_{q\eta}$ are then given by
\begin{subequations}\label{eq: lambda_tilde_pm}
\begin{align}
\tilde{\lambda}_{q\eta}^{(+)}&=\gamma_q\,,\\
\intertext{which is independent of $\eta$, and }
\tilde{\lambda}_{q\eta}^{(-)}&=\begin{cases}
+\gamma_q\,,&\text{ for }\eta=1\\-\gamma_q\,,&\text{ for }\eta=2\,. \end{cases}
\end{align}
\end{subequations}

To see that the effect of the leads on the impurity is indeed the same in the local and non-local schemes, we note that level-doubling replaces the original hybridization function, given by Eq.~\eqref{eq: definition Delta}, by
\begin{align}
\label{eq: new definition Delta}
\tilde{\Delta}^{R/K}_{ij,\alpha}(\omega)&=\sum_{k\eta}\tilde{v}^\pstar_{iq\eta}\tilde{v}^*_{jq\eta}\,\tilde{g}^{R/K}_{q\eta,q\eta}(\omega)\, ,
\end{align}
where the correlators $\tilde{g}^{R/K}_{q\eta,q\eta}(\omega)$ are given by Eq.\,\eqref{eq: g Lindblad lambda} with $q$ replaced by $\{q \eta\}$ and $\lambda^{(\pm)}$ by $\tilde{\lambda}^{(\pm)}$.  \Eq{eq: new definition Delta} yields expressions identically equal to the original hybridization~\eqref{eq: definition Delta}. For the retarded component this follows from 
\begin{subequations}
\begin{align}
\sum_\eta \tilde{v}_{iq\eta}\tilde{v}^*_{jq\eta}&=v_{iq}v^*_{jq}\, . \\
\intertext{Similarly, the Keldysh component is the same as the original one since}
\sum_\eta \tilde{\lambda}^{(-)}_{q\eta}\tilde{v}_{iq\eta}\tilde{v}^*_{jq\eta}&=\lambda^{(-)}_{q}v_{iq}v^*_{jq}\, .
\label{eq:shift-T-V-dependence}
\end{align}
\end{subequations}
The last equation explicitly shows how, when passing from the non-local to the local setup, the Fermi-function information encoded in the Lindblad rates $\lambda_{q}^{(-)}$ of the former is shifted into the couplings $\tilde{v}_{iq\eta}$ of the latter. This is illustrated schematically in Fig.~\ref{fig: Lead Levels}(b).

For a uniform discretization in energy space, the rates $\gamma_q$ can be chosen independent of $q$. Hence, the level-doubled Lindblad matrices $\tilde{\Lambda}^{(1,2)}$ for each $\eta=1,2$ are separately proportional to the identity. Thus, they will remain so under the linear transformations used to map impurity models to quantum chains, provided that these transformations do not mix the two `channels' $\eta=1$ and $\eta=2$. We have thus found what we were looking for: an LDDL scheme reproducing the correct hybridization with Lindblad driving terms that will remain local when the leads are represented in terms of chains.

\begin{figure}
\includegraphics[width=\linewidth]{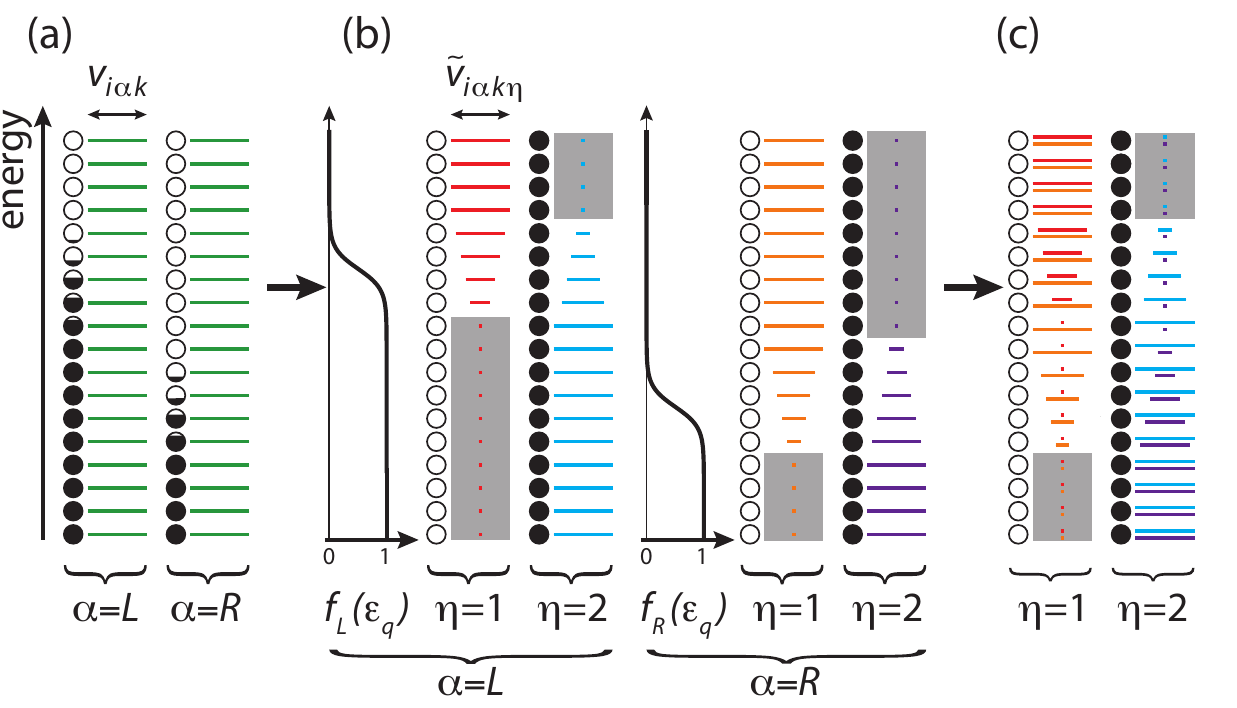}
\caption{Schematic depiction of the level-doubling construction scheme for two leads, $\alpha=L$ and $R$, assuming constant values of $v_{i \alpha k}$. (a) Original levels of the left and right leads, described by $c_{\alpha k}^{(\dagger)}$ operators.  (b) After level doubling, each lead $\alpha$ is represented by two sets of auxiliary levels, distinguished by $\eta = 1$ and 2 and Lindblad-driven towards occupancy 0 and 1, respectively. These levels are described by $\tilde{c}_{\alpha k\eta}^{(\dagger)}$ operators, whose coupling strengths $\tilde{v}_{i \alpha k \eta}$ (indicated by the width of the horizontal lines) depend on the Fermi function $f_\alpha(\varepsilon_q)$ of that lead (depicted by smooth black curves).  For $\eta=1$ (or 2) all those auxiliary levels decouple for which $f_\alpha(\varepsilon_q)\approx 1$ (or 0), indicated by grey shading. (c) For a model involving just a single impurity level, only certain linear combinations of $L$ and $R$ auxiliary lead operators, the $\tilde{b}_{k\eta}^{(\dagger)}$ operators of Eq.~\eqref{eq: combine Lketa Rketa (a)}, couple to the impurity; they are depicted here by double lines representing the couplings $\tilde v_{Lk \eta}$ and $\tilde v_{Rk\eta}$, with grey shading indicating vanishing couplings. }
\label{fig: Lead Levels}
\end{figure}

At first glance, the local setup comes at a high price, namely twice as many lead levels as before, due to the additional label $\eta$. This, however, is not the full truth: for all levels with energies $|\varepsilon_{k}-\mu_\alpha|\gg T_\alpha$, the value of the Fermi function $f_\alpha\left(\varepsilon_q\right)$ will be either one or zero. Therefore, by Eq.~(\ref{eq: Vk_tilde}) either $\tilde v_{i q,\eta= 1}$ or $\tilde v_{i q,\eta= 2}$ will vanish, implying that one of the two corresponding auxiliary modes, with either $\eta = 1$ or 2, will decouple from the impurity [indicated by grey shading in Fig.~\ref{fig: Lead Levels}(b)]. Thus, the number of impurity-coupled auxiliary levels in each lead is actually equal to the number of original levels throughout the energy ranges where the Fermi function equals 1 or 0, and twice that number only in the intermediate range that encompasses the step in $f_\alpha (\varepsilon_k)$. In particular, for $T\to 0$, this intermediate range shrinks to zero.

Moreover, the local setup results in a further major simplification stemming from the fact that its Lindblad rates $\tilde \lambda^{(1,2)}$ are independent of $\alpha$: depending on the exact form of the impurity and the coupling to the impurity, certain linear combinations of auxiliary modes from \textit{different} leads may decouple. We illustrate this for the case of two spinless leads $\displaystyle{\alpha=\{L,R\}}$ coupled to one spinless impurity level, using the same discretization for the two leads, $\varepsilon_{\alpha k}=\varepsilon_{k}$. For such a model, the index $i=1=d$ can be dropped in the coupling matrix elements. Hence, we can combine the auxiliary modes $\{L k\eta\}$ and $\{R k\eta\}$ by defining
\begin{subequations}\label{eq: combine Lketa Rketa}
\begin{align}
\label{eq: combine Lketa Rketa (a)}\tilde{b}_{k\eta}=\frac{1}{\sqrt{\sum_{\alpha}|\tilde{v}_{\alpha k \eta}|^2}}\left(\tilde{v}_{L k\eta}\tilde{c}_{L k\eta} + \tilde{v}_{R k\eta}\tilde{c}_{R k\eta}\right)\,,\\
\tilde{b}'_{k\eta}=\frac{1}{\sqrt{\sum_{\alpha}|\tilde{v}_{\alpha k \eta}|^2}}\left(\tilde{v}_{R k\eta}\tilde{c}_{L k\eta} - \tilde{v}_{L k\eta}\tilde{c}_{R k\eta}\right)\,.
\end{align} 
\end{subequations}
Only the $\tilde b_{k\eta}$ modes couple to the impurity, whereas the $\tilde{b}'_{ k\eta}$ modes do not. This is completely analogous to what is done for such models in equilibrium calculations.  In nonequilibrium, however, where $f_L \neq f_R $, such a transformation would not have been useful if performed in the original non-local setup, because the original Lindblad rates $\lambda^{(1,2)}$ actually depend on $f_\alpha$, so that transforming them using (\ref{eq: combine Lketa Rketa}) would generate a coupling between the modes $\tilde{b}_{k\eta}$ and $\tilde{b}^\prime_{k\eta}$ via the dissipative Lindblad terms.  In the local setup, however, where the $\tilde \lambda^{(1,2)}$ are independent of $\alpha$, no such coupling is generated, so that the $\tilde{b}'_{ k\eta}$ modes decouple altogether. We are thus left with only two impurity-coupled auxiliary channels, with modes $\tilde b_{k1}$ and $\tilde b_{k2}$, but they have a completely different interpretation than the two physical leads from which we started, with modes $c_{kL}$ and $c_{kR}$. This is illustrated in Fig.~\ref{fig: Lead Levels}(c): it depicts the linear combinations $\tilde{b}_{k\eta}$ in Eq.~\eqref{eq: combine Lketa Rketa} that couple to the impurity using double lines. The modes $\tilde{b}_{k\eta}'$ are omitted as they decouple from the model.

Figs.~\ref{fig: Lead Levels}(c) and ~\ref{fig: Lead Levels}(a) together nicely summarize the level count of impurity-coupled auxiliary versus original levels. Within the dynamical window, defined by the energy range in which $f_L(\varepsilon_q) \neq f_R(\varepsilon_q)$, the number of impurity-coupled auxiliary lead levels in the local setup [Figs.~\ref{fig: Lead Levels}(c)] is the same as the number of physical lead levels in the original non-local setup [Figs.~\ref{fig: Lead Levels}(a)], corresponding to a full two-channel calculation. Outside the dynamical window, where $f_L(\varepsilon_q)=f_R(\varepsilon_q)=1$ (or 0), the auxiliary levels corresponding to $\eta=1$ (or 2) decouple from the impurity (as indicated by grey shading), hence here the number of impurity-coupled auxiliary levels equals half the number of original levels. This reduction of levels is easily understood considering that outside the dynamical window we effectively have an equilibrium situation (in that $f_L(\varepsilon_q)=f_R(\varepsilon_q)$ there) and can therefore use the same decoupling transformation as that used routinely in equilibrium calculations. Note also that in the special case of $T=0$, the modes $\tilde{b}_{k\eta}$ within the dynamical window are identical to either $\tilde{c}_{Lk\eta}$ or $\tilde{c}_{Rk\eta}$.

Of course, such a decoupling of modes is not guaranteed to occur in general for multi-level models. For example, it does not happen for a model with more than one impurity level where each impurity level couples differently to the leads.

The operators from the original non-local and new local setups, $c_{\alpha k}$ and $\tilde{c}_{\alpha k \eta}$, are obviously not related by any unitary transformation (after all, they even differ in number). Expressions for the currents into the leads $\alpha$ therefore have to be found using the new Lindblad equation in the local chain representation. Given the fact that the lead index $\alpha$ is still a well-defined quantity, this can straightforwardly be done by evaluating $e\dot{N}_d=0=\sum_\alpha J_\alpha$ analogously to Sec.~\ref{sec: current RLM}, resulting in expressions analogous to Eq.~\eqref{eq: current, Greensfunctions}, with $q\to q\eta$ and $\sum_k\to \sum_{k\eta}$.  For the above example of one spinless local mode coupled to two spinless leads, the expectation values $\braket{\tilde{c}_{\alpha k\eta}^\dagger c_{d}^\pdag}_\text{NESS}$ needed for the evaluation of the current can then be expressed in terms of $\braket{\tilde{b}_{ k\eta}^\dagger c_{d}^\pdag}_\text{NESS}$:
 \begin{align}
\braket{\tilde{c}_{\alpha k\eta}^\dagger c_{d}}_\text{NESS}=\frac{1}{\sqrt{\sum_{\alpha}|\tilde{v}_{\alpha k \eta}|^2}}\tilde{v}_{\alpha k\eta}\braket{\tilde{b}_{k \eta}^\dagger c_d}_\text{NESS}\,,
\end{align}
where we used the fact that the mode $\tilde{b}'_{ k \eta}$ decouples from the impurity level, $\braket{\tilde{b}^{\prime\dagger}_{ k \eta} c_{d}^\pdag}_\text{NESS}=0$.

For the RLM it is straightforward to verify that Eqs.\ (\ref{eq: GR}) and (\ref{eq: GK}) yield the same results for $G^R_{dd}(\omega)$ and $G^K_{dd}(\omega)$ when evaluated within the local setup as in the original non-local setup [Eqs. (\ref{eq: GR_Lindblad RLM}) and (\ref{eq: GK_Lindblad RLM})]. Analogously, also the results for the current (\ref{eq: current_Lindblad}) and the occupation of the local level (\ref{eq: DotOccupation_Lindblad}) can easily be reproduced.

Let us note that this concept of representing thermal leads by ``holes'' and ``particles'' with couplings that depend on the Fermi function has also be found using the thermofield approach \cite{PhysRevA.92.052116}.

\section{Log-Linear Discretization}\label{sec: Logarithmic Discretization}
\begin{figure}[b]
\centering
\includegraphics[width=0.45\textwidth]{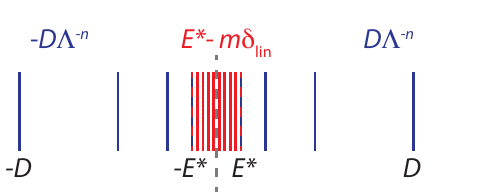}
\caption{Sketch of the suggested discretization: the high-energy intervals are discretized logarithmically, while a window $[-E^*,E^*]$ large enough compared to the dynamical window is discretized linearly.}
\label{fig: discretization}
\end{figure}

In quantum impurity models it is often of great interest to consider a wide range of different energy scales, e.\,g.\ for models exhibiting Kondo physics. Within the numerical renormalization group, one therefore uses a logarithmic discretization, $\varepsilon_k\sim \pm D \Lambda^{-k}$ with $\Lambda>1$. This leads to a very efficient description of the renormalization of impurity properties, since much fewer discrete levels are needed to reach low energy scales than when discretizing linearly. For such a logarithmic discretization it is necessary to have an explicit energy reference, the physics around which is resolved in greater detail. In equilibrium, this reference point is defined by the chemical potential. In contrast, in situations of steady-state nonequilibrium, there is not one single Fermi edge, but a dynamical window that needs to be described accurately, defined by the energy range contributing to transport. Within this window a logarithmic discretization does not seem to be appropriate. Therefore a more flexible discretization scheme is desirable \cite{PhysRevB.75.241103,0295-5075-73-2-246,PhysRevB.87.115115}. Here, we advocate the use of a "log-linear" discretization scheme which is linear within a window sufficiently large compared to the dynamical window and logarithmic for energies outside this range, similar to the approach used in Ref.\ [\onlinecite{PhysRevB.80.165117}]. The underlying rationale is that within the dynamical window there is no energy scale separation. Therefore, the discretization should not introduce any artificial structure to the calculation, and thus be uniform. Here, we assume a symmetric setup and therefore a symmetric range $[-E^*,E^*]$ that is discretized linearly with level spacing $\delta_\text{lin}$, as depicted in Fig.~\ref{fig: discretization}.

\begin{figure*}
\includegraphics[width=.32\linewidth]{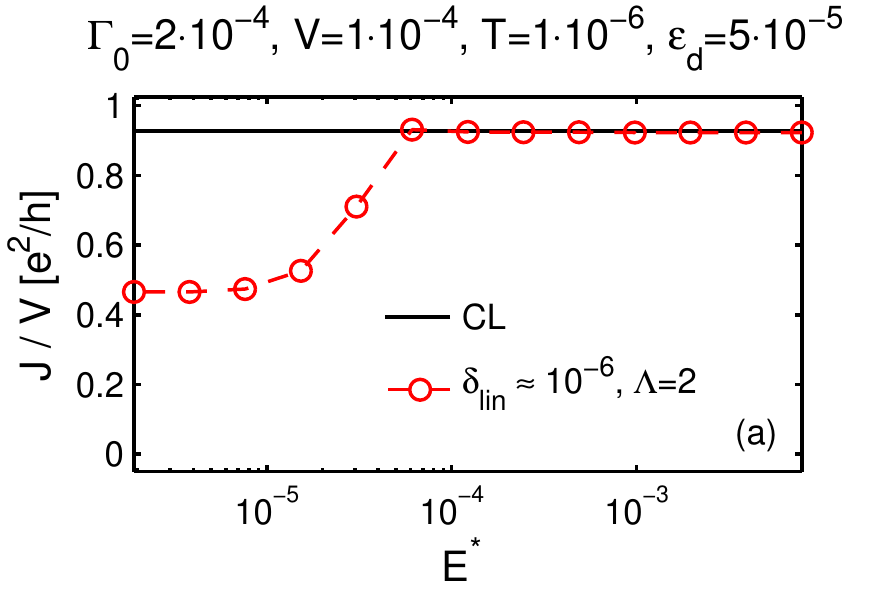}
\includegraphics[width=0.32\linewidth]{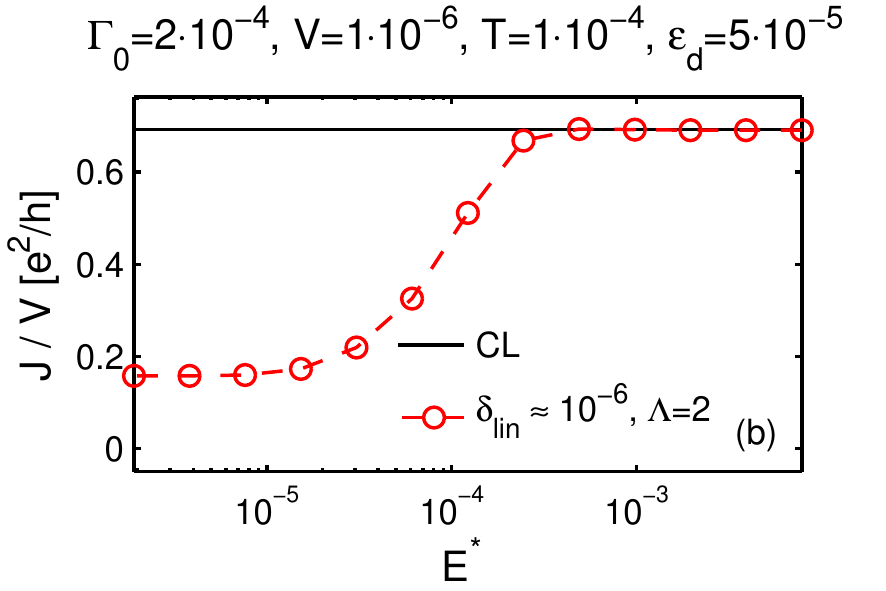}
\includegraphics[width=0.32\linewidth]{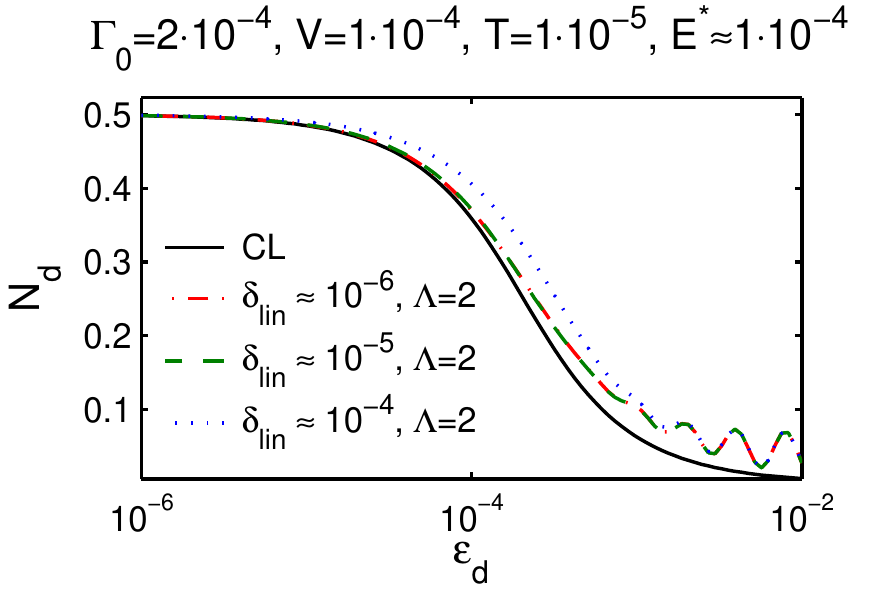}
\caption{(a) and (b) show the current through the local level of the RLM as given by Eq.\,(\ref{eq: current_Lindblad}) using a discretization, which is linear within the dynamical window $[-E^*,E^*]$ and logarithmic outside as a function of $E^*$. In (a), the voltage is large compared to temperature and the correct value for the current can only be obtained if $E^*\gtrsim 1.2\mu_L=-1.2\mu_R$\,. In (b), temperature is larger than voltage and $E^*\gtrsim 4T$ is needed. In (c), the occupation of the local level in the RLM as given by Eq.\,(\ref{eq: DotOccupation_Lindblad}) is shown as a function of the level position $\varepsilon_d$, again using the log-linear discretization. For $\varepsilon_d\gtrsim E^*$ we see deviations from the CL result, see Appendix~\ref{sec: appendix A} for details. For all three panels, $\Lambda=2$ and we used $\gamma_q=\gamma=\delta_\text{lin}$ for the linear and the logarithmic states. The number of lead levels is approximately given by $M_\text{lin}={2E^*}/{\delta_\text{lin}}$ plus $M_\text{log}=-2\text{log}(E^*)/\text{log}(\Lambda)$.}
\label{fig: logarithmic}
\end{figure*}

We have argued above that the strength of the Lindblad driving $\gamma_q$ for a given lead level should be comparable to or larger than the width $\delta_q$ of the corresponding energy interval. Furthermore, $\gamma_q$ needs to be $q$-\textit{in}dependent to permit the mapping onto a local chain that we suggested in Sec.~\ref{sec: Local chain}. This seems to be incompatible with the logarithmic discretization scheme, since the latter features energy intervals whose widths depend on $q$. Note, though, that the logarithmically discretized regime by construction describes excitations on energy scales much larger than the energy scales on which transport takes place. These excitations are not affected by nonequilibrium physics but are only involved in renormalization effects, which (as we know from the success of NRG) are well described even if these levels are not broadened at all. In other words, the condition $\delta_q \simeq \gamma_q$ is not needed for energy scales far outside the transport window, but only for levels that are involved in dissipative effects.  We may thus use a Lindblad driving $\gamma_q=\gamma=\delta_\text{lin}$ for the logarithmically discretized states as well, although this is much smaller than the widths of the corresponding energy intervals. Note that this implies that, if one solves the Lindblad equation numerically using time evolution or some optimization scheme, the starting state should be chosen close enough to the steady state (which for the high-energy states means low enough in energy), because high-energy modes are barely damped. Also, as mentioned earlier, the Lindblad driving does not need to broaden the peak structure arising from the discretization. If needed, this broadening of the discrete peak structure can be done by hand after solving the Lindblad equation, analogously to the broadening in equilibrium NRG calculations \cite{PhysRevLett.99.076402,RevModPhys.80.395}.

Below, we will discuss the implications of the choice $\gamma_q=\gamma=\delta_\text{lin}$ within the RLM, bearing in mind a caveat: for the RLM the nonequilibrium Green's function $G_{dd}^R(\omega)$ is equal to its equilibrium pendant, which is not true for general interacting impurity models. Therefore, the RLM does not allow a fully general check whether the choice $\gamma_q=\gamma=\delta_\text{lin}$ is able to capture all nonequilibrium properties of the high-energy states occurring in this Green's function. This will be left for future studies.
 
In Figs.~\ref{fig: logarithmic}(a) and \ref{fig: logarithmic}(b) we plot the current for the spinless RLM as given in Eq.\,(\ref{eq: current_Lindblad}) again using the symmetric setup defined in (\ref{eq: Gamma=box two channels}) with the same discretization for both leads. Here, however, we use the suggested discretization with energy intervals $[\pm\Lambda^{-(n-1)},\pm\Lambda^{-n}]$ for $n=1\dots N_\text{log}$, where $N_\text{log}$ is defined by $\Lambda^{-N_\text{log}}=E^*$. The window $[-E^*,E^*]$ is discretized linearly using $2N_\text{lin}$ energy intervals of size $\delta_\text{lin}=E^*/N_\text{lin}$. For the prefactor of the Lindblad driving we use $\gamma_q=\gamma=\delta_\text{lin}$ for both the logarithmically and the linearly discretized energy intervals. The current is plotted for different values of $N_\text{log}$ corresponding to different values of $E^*$. The level spacing is kept approximately the same, which means that more levels are needed for larger $E^*$. Evidently, if $E^*$ is large enough and $\delta_\text{lin}$ small enough, it is possible to reproduce the value for the current that one obtains in calculations using continuous thermal leads. Furthermore, the two plots illustrate which energy range should be resolved linearly: In the first panel we have $T\ll V$. Here, the dynamical window is defined by the two chemical potentials and the full current is only recovered if $E^*\gtrsim 1.2\mu_L=-1.2\mu_R$. In the second panel temperature becomes the relevant energy scale due to $T\gg V$, the two Fermi functions differ in an energy range defined by temperature, and therefore $E^*\gtrsim4\cdot T$ is needed.  

Figure \ref{fig: logarithmic}(c) shows the occupation number of the local level given in (\ref{eq: DotOccupation_Lindblad}) as function of the level position $\varepsilon_d$. Only positive values of $\varepsilon_d$ are considered. The occupation for negative level position $\varepsilon_d$ can be deduced from this data by $N_d(-\varepsilon_d)=1-N_d(\varepsilon_d)$. This relation can be shown both for the Lindblad result (\ref{eq: DotOccupation_Lindblad}) as well as for the result of continuous leads (\ref{eq: DotOccupation_Keldysh}). Here, the suggested discretization only works well for $\varepsilon_d\lesssim E^*$. For $\varepsilon_d\gtrsim E^*$, the Lindblad result for the occupation number deviates from the value obtained for continuous leads. This deviation is independent of $\delta_\text{lin}$ and shows oscillations that correspond to the logarithmically discretized lead levels. This indicates that the error stems from the logarithmically discretized part of the lead.

At first glance, it is not surprising that an error arises when $\varepsilon_d$ becomes so large that it falls within the logarithmic discretized part of the spectrum. In this case, the energy range around $\varepsilon_d$ where the relevant physics takes place is not sufficiently resolved. Note though that for large $\varepsilon_d$ standard NRG calculations using a logarithmic discretization for the full energy range are able to determine the equilibrium occupation number with a much higher accuracy than the LDDL approach with log-linear discretization. Therefore, a detailed analysis of how this error comes about and how its effects can be minimized is offered in Appendix~\ref{sec: appendix A}. 

Let us finally comment on the use of the numerical renormalization group within the LDDL setup. Applying the mapping onto a local chain as described in Sec.\,\ref{sec: Local chain}, the hoppings corresponding to the logarithmically discretized energy range will fall off exponentially, as for standard NRG Wilson chains\cite{RevModPhys.47.773,RevModPhys.80.395}. Thus, it should be possible to construct an effective many-body basis for this part of the chain using NRG \cite{PhysRevLett.95.196801,PhysRevB.74.245113}. Assuming that the nonequilibrium at low energy scales does not affect the high-energy physics, standard NRG truncation of this basis is justified. For the treatment of the linearly discretized dynamical window there is no energy-scale separation and other MPS techniques such as tDMRG \cite{1742-5468-2004-04-P04005,PhysRevLett.93.076401,PhysRevLett.93.040502} have to be used. This approach is in close analogy to the hybrid NRG-DMRG approach of Ref.~[\onlinecite{PhysRevB.87.115115}].

\section{Conclusion and Outlook}
\label{sec:conclusion-outlook}

In summary, we have explored the suitability of Lindblad-driven discretized leads for the description of nonequilibrium steady-state physics in models in which a correlated impurity is coupled to non-interacting leads and each lead is independently held at a fixed chemical potential and temperature. For quadratic models governed by Lindbladian dynamics we have introduced a simple approach to calculate steady-state Green's functions. We have shown that the additional Lindblad reservoirs introduce a broadening for the discretized lead levels and that the Lindblad rates can be tuned to provide an exact representation of thermal reservoirs in the continuum limit. The approach, therefore, is appropriate for the description of steady-state nonequilibrium of arbitrary impurities of the kind that arises due to an applied voltage or temperature difference. For the quadratic resonant level with applied voltage, we analytically calculated the current through the local level and the occupation of the local level within the Lindblad setup and found perfect agreement with the results that one obtains using standard calculations for continuous thermal leads. 

To explore heat current due to an applied temperature difference, one could study how the energies of the leads change due to their coupling to the impurity, starting from $\dot{H}_{L/R}$ to define left and right energy currents, in a manner similar to the definitions used here for the charge current. 

Finally, we presented first steps towards a future numerical determination of the steady state using MPS/MPO methods, showing how the leads can be represented in terms of chains with the desirable property that both the Lindblad driving terms and the Hamiltonian dynamics are local. We also advocated the use of a log-linear discretization scheme in this context that should permit the exploration of exponentially small energy scales.

Our analysis shows that the LDDL approach constitutes a promising starting point for a systematic treatment of quantum impurity models in steady-state nonequilibrium using MPS/MPO-based numerical approaches. Future work will have to explore which of these approaches targeting the steady-state solution of the Lindblad equation turns out to be the most efficient.

\acknowledgements
We  acknowledge fruitful discussions with I.~Weymann, T.~Prosen and H.~Kim. This work was supported by the German-Israeli-Foundation through I-1259-303.10. F.\,S., A.\,W. and J.\,v.\,D. were also supported by the Deutsche Forschungsgemeinschaft through SFB631, SFB-TR12, and NIM. A.\,W. was also supported by WE4819/1-1 and WE4819/2-1. M.\,G. was also supported by the Israel Science Foundation (Grant 227/15) and the US-Israel Binational Science Foundation (Grant 2014262). A. D. and E. A. were supported by the Austrian Science Fund (FWF): P24081 and P26508, and by NAWI Graz.

\appendix

\section{Detailed error analysis for the log-linear discretization discussed in Section~\ref{sec: Logarithmic Discretization}}\label{sec: appendix A}
In Section \ref{sec: Logarithmic Discretization} we have seen that for a discretization that is logarithmic for high energies and linear within the dynamical window, the occupation of the local level in the RLM calculated using the LDDL scheme deviates from the exact continuum result. This error appears if the position of the local level $\varepsilon_d$ lies within the logarithmically discretized energy range. Moreover, this error is independent of $\delta_\text{lin}$ and shows oscillations that correspond to the logarithmically discretized lead levels. 

To understand where this deviation comes from, we divide the integrand in Eq.\,(\ref{eq: DotOccupation_Lindblad})  into two parts, $|G^R_{dd;\text{DL}}(\omega)|^2$ and $\sum_qf_\alpha(\varepsilon_k)|v_q|^2\,\Lorentz{\gamma_q}{k}$. These functions have to be compared to $|G^R_{dd;\text{CL}}(\omega)|^2$ and $\sum_\alpha f_\alpha(\omega)\Gamma_{\alpha;\text{CL}}(\omega)$ in Eq.\,(\ref{eq: DotOccupation_Keldysh}). Assume now that \mbox{$\varepsilon_d\gg E^*$}. In this case $G_{dd;\text{DL}}^R(\omega)$ is non-zero mainly for $\omega>E^*$. In this $\omega$ region the sum over $q$ consists of tails of Lorentz peaks stemming from the lead levels with $\varepsilon_k$ below or within the dynamical window only, while the contribution of all other levels is exponentially suppressed by $f_\alpha(\varepsilon_k)\approx0$. Hence, for \mbox{$\omega>E^*$}, the sum over $q$ in Eq.\,(\ref{eq: DotOccupation_Lindblad}) is \textit{polynomially} suppressed by the small peak width $\gamma$, whereas the corresponding expression for continuous leads in (\ref{eq: DotOccupation_Keldysh}) is \textit{exponentially} suppressed by the Fermi functions $f_\alpha(\omega)$\,. The small but finite overlap of the Lorentz tails with the function $|G_{dd;\text{DL}}^R(\omega)|^2$ in (\ref{eq: DotOccupation_Lindblad}), which does not exist in the exact formula (\ref{eq: DotOccupation_Keldysh}), is the explanation for the deviation of the Lindblad result from the CL value.

But why is this error independent of $\gamma=\delta_\text{lin}$, although the Lorentz tails obviously scale with $\gamma$? The answer lies in the peak structure of $|G^R_{dd;\text{DL}}(\omega)|^2$: For $|\omega|> E^*$, the lead is logarithmically discretized and the Lindblad broadening $\gamma=\delta_\text{lin}$ is small compared to the size of the underlying energy intervals. Therefore, $|G^R_{dd;\text{DL}}(\omega)|^2$ contains sharp peaks in this $\omega$ region and the peak widths scale with $\gamma$. However, because $G_{dd;\text{DL}}^R(\omega)$ is a physical Green's function, the area beneath the real and imaginary parts of this function is represented correctly and therefore independent of $\gamma$. Assuming that the peaks are well separated, this implies, that the integral over $|G_{dd;\text{DL}}^R(\omega)|^2$ scales approximately with $\gamma^{-1}$ in this logarithmically discretized region. Decreasing $\delta_\text{lin}=\gamma$, therefore, does not reduce the error in the occupation number, because, while the sum over the tails of the Lorentz functions scales with $\gamma$, the area of $|G^R_{dd;\text{DL}}(\omega)|^2$ scales with ${\gamma}^{-1}$, leaving the total error approximately the same.

In contrast, if $\varepsilon_d$ lies within the dynamical window, the main contribution of $|G^R_{dd;\text{DL}}(\omega)|^2$ (and therefore the main contribution of the integrand) lies within the linearly discretized window. Here, the peaks of $|G_{dd;\text{DL}}^R(\omega)|^2$ strongly overlap and therefore the integral over $|G_{dd;\text{DL}}^R(\omega)|^2$ is $\gamma$-independent. In other words, the integrand is represented as a smooth function within the linearly discretized window. Hence, if $\delta_\text{lin}=\gamma$ is small enough to resolve all relevant features, the integrand coincides with the exact CL integrand and no error is observed.

The occupation of the local level for negative $\varepsilon_d$ can be deduced by $N_d(-\varepsilon_d)=1-N_d(\varepsilon_d)$. Therefore, for $\varepsilon_d\ll-E^*$ an error analogous to that for $\varepsilon_d\gg E^*$ occurs.

One possibility to avoid the error is to replace the sum over $q$ by its continuum counterpart: $\sum_qf_\alpha(\varepsilon_k)|v_q|^2\,\pi\Lorentz{\gamma_q}{k}\to\sum_\alpha f_\alpha(\omega)\Gamma_{\alpha;\text{CL}}(\omega)$. This is equivalent to using the standard form of the occupation number given by the continuum limit of Eq.\,(\ref{eq: DotOccupation_Keldysh spectral function}) but with the exact Green's function replaced by the Green's function deduced from Lindblad formalism. In general, i.e.\ also for interacting models, which cannot be solved analytically, this procedure corresponds to deducing only the Green's function from the Lindblad approach and then calculating the occupation number using standard Green's function techniques. (Note, though, that numerically evaluating $G_{dd;\text{DL}}^R(\omega)$ can be computationally more demanding than simply evaluating expectation values. For example, this is the case in the quantum trajectory approach.)

Why does the error not occur for a linear discretization? In fact, it does, but can be scaled down using more lead levels. When discretizing the full bandwidth $[-D,D]$ linearly, $|G_{dd;\text{DL}}^R(\omega)|^2$ is represented by a smooth function within the full band, because the Lindblad broadening is comparable to the size of the energy intervals everywhere. The area beneath $|G_{dd;\text{DL}}^R(\omega)|^2$, therefore, does not depend on $\gamma$, while the contribution of the Lorentz tails for large $\omega$ can be reduced using a smaller value of $\delta=\gamma$. (Note that the number of lead levels $q$ that we sum over, scales with $\delta^{-1}\sim\gamma^{-1}$. However, this $\gamma$-dependence is canceled by the $\gamma$-dependence of $|v_q|^2$ which scales with $\delta\sim\gamma$. Therefore, the scaling of  $\sum_qf_\alpha(\varepsilon_k)|v_q|^2\,\pi\Lorentz{\gamma_q}{k}$ with $\gamma$ stemming from the Lorentz tails is preserved.) Nonetheless, also for a linear discretization, it could be advisable to replace the sum over $q$ by its continuum representation as described above to reduce the error for a fixed number of states. 

Another question arising immediately is why this kind of error is not visible in the current. If we look at Eq.\,(\ref{eq: current_Lindblad}) we find two major differences compared to the analysis of the occupation number above. First, the sum over the lead levels $\sum_k |v_k|^2\,\pi\Lorentz{\gamma_k}{k}\left(f_L(\varepsilon_k)-f_R(\varepsilon_k)\right)$ contains the difference of Fermi functions instead of a sum. This implies that only the lead levels corresponding to the linearly discretized dynamical window contribute, while the contribution of the logarithmically discretized intervals is exponentially suppressed. Nevertheless, the tails of the Lorentz peaks in this sum leak out to high values of $|\omega|$, whereas in the formula for continuous thermal leads contributions from this $\omega$ range are exponentially suppressed. The second and relevant difference is the fact that, while the sum over $k$ is multiplied by $|G_{dd;\text{DL}}^R(\omega)|^2$ in the formula for the occupation of the local level, it is multiplied by $\text{Im}(G_{dd;\text{DL}}^R(\omega))$ in the formula for the current. Both functions are strongly peaked in the logarithmically discretized region, but as explained above, the integral over $\text{Im}(G_{dd;\text{DL}}^R(\omega))$ is independent of $\gamma$, whereas the integral over $|G_{dd;\text{DL}}^R(\omega)|^2$ scales with $\gamma^{-1}$. Due to this difference the error in the occupation number is independent of $\gamma$ while the error in the current is proportional to $\gamma$ and can therefore be reduced using smaller $\delta_\text{lin}=\gamma$. But again, for fixed $\delta_\text{lin}$,  it could be possible to reduce the error of the Lindblad result by using the continuum analog of the sum over $k$, analogously to what was described for the occupation number above.

\section{Quantum regression theorem  for Fermion operators}\label{sec: appendix B}

\newcommand{\fs}{s}  
\newcommand{\fr}{r}  
\newcommand{\FS}{S}  
\newcommand{\FRR}{R}  
\newcommand{\FFR}{{\hhh R}}  
\newcommand{\NR}{{N_\FRR}}  
\newcommand{\NS}{{N_\FS}}
\newcommand{\fu}{{\rm full }} 
\newcommand{\hsr}{H_{S,R}}
\newcommand{\hhfs}{S}  
\newcommand{\al}{\alpha}
\newcommand{\beq}{\begin{equation}}
\newcommand{\eeq}{\end{equation}}
\def\beqa#1\eeqa{\begin{align}#1\end{align}}
\newcommand{\pcite}[1]{{\bf[CITE: #1]}}
\newcommand{\pref}[1]{{\bf[REF OF PAPER: #1]}}
\newcommand{\tr}{\mbox{tr}}
\newcommand{\hha}{\hhh {\rhotilde}}
\newcommand{\li}{{\cal L}}
\newcommand{\ch}{{\cal H}}
\newcommand{\id}{{\mathbb{I}}}
\newcommand{\hhh}[1]{\hat{#1}}
\newcommand{\hhhNR}{\hat{N}_{\FRR}}
\newcommand{\hhhNS}{\hat{N}_{\FS}}

In this appendix, we derive the Lindblad equation Eq.\,(\ref{eq: rhotilde_dot}) for $\rhotilde_{\C}(t)$, in which the operator $\C$ from Eq.~\eqref{eq: tilde rho def} contains an odd number of fermionic operators. 
It is an extension of the so-called quantum regression
  theorem (QRT) \cite{Breuer2002,Gardiner2000,carmichael1}
to the case of fermionic operators \cite{Schaller2014}.

\subsection{Time evolution of reduced density matrix}

We start by showing that in the fermionic case the density matrix itself obeys the same Lindblad equation (\ref{eq: Lindblad quadratic}) as for bosons. The usual derivation of the Lindblad equation within the Born-Markov approximation (BMA)~\cite{Breuer2002,Gardiner2000,carmichael1}$^{,}$\footnote{The same derivation applies for the so-called singular-coupling limit \cite{Breuer2002,Gardiner2000,carmichael1}} starts from a system-reservoir Hamiltonian in the form of a sum of tensor products of operators acting on the system and reservoir separately. For the fermionic case, however, one generally has a system-reservoir Hamiltonian of the form \beq
\label{hsr0}
\hsr = \sum_{\al} \fr_\al \fs_\al + {\rm h.c.} ,  \eeq where $\fr_\al$ ($\fs_\al$) are
reservoir (system) operators containing an odd number of fermionic
operators, i.e.  \beq \{\fs_\al,\fr_\beta\} = 0 \;.  \eeq Since the
operators $\fr_\al$ and $\fs_\al$ \textit{anti}commute, \eqref{hsr0}
cannot be interpreted as a tensor product between operators
acting independently on the reservoir ($\ch_{\FRR}$) and system
($\ch_{\FS}$) Hilbert spaces.  For the sake of clarity, in the present
Appendix it will be convenient to
distinguish between when a particular operator, such as, e.\,g. $\fs_\al$ acts
on the reservoir-system product Hilbert space $\ch_{\FRR\FS}$ or just
on one of the two separate spaces.  In the latter case, we will add a
hat (``$\hhh{~}$'') to the operator. (In the main
  text we do not use hats because there nearly all operators act on
  the system's Hilbert space and the few exceptions can easily be
  recognized from the context.)  For definiteness, we adopt the
convention that product states in $\ch_{\FRR\FS}$ are understood in
the following order: \beq \ket{\FRR}\otimes\ket{\FS} \, , \eeq where
$\ket{\FRR}\in \ch_\FRR$ and $\ket{\FS}\in \ch_\FS$.  Due to the
properties of fermion operators we, thus, have the relation \beqa
\fs_\al &=  (-1)^{\hhhNR} \otimes \hhh \fs_\al\,, \\
\fr_\al &= \hhh \fr_\al \otimes \hhh{\id}_\FS\,, \eeqa where
${\NR}=\hhhNR\otimes\hhh{\id}_\FS$ is the operator counting the number
of fermions in the reservoir.  With this notation, \eqref{hsr0} can be
written in tensor form as
\beq
\label{hsr}
\hsr = 
\sum_{\al} \FFR_\al\otimes \hhh \fs_\al \; .  
\eeq 
where we have introduced
\beq
\FFR_\al \equiv \hhh r_\alpha (-1)^{\hhhNR} \;.
\eeq

In this form, it \textit{is} possible to directly apply the standard BMA derivation
of the Lindblad equation~\cite{Breuer2002,Gardiner2000,carmichael1}.  According to
that derivation, the expression for its coefficients depend on the 
Fourier transforms of the
unperturbed reservoir correlation
functions~\cite{Breuer2002,Gardiner2000,carmichael1} \beq
\label{cab}
C_{\al,\beta}(t) = \tr_\FRR \left(\FFR^\dag_\al(t) \FFR_\beta \hhh{\rho}_\FRR\right)\,,
\eeq
with $\FFR_\alpha(t)=\e^{\ii\hat{H}_Rt}\FFR_\alpha\e^{-\ii\hat{H}_Rt}$.
The only requirement is that one starts with a reservoir-system Hamiltonian in the form of a tensor product.
These correlation functions can be rewritten as
\begin{align}
\notag C_{\al,\beta}(t) &= \tr_\FRR \left( (-1)^{\hhhNR(t)}\ \hhh \fr^\dag_\al(t) \hhh \fr_\beta\  (-1)^{\hhhNR} \hhh{\rho}_\FRR\right)\\
&= \tr_\FRR\left(   \hhh \fr^\dag_\al(t) \hhh \fr_\beta \hhh{\rho}_\FRR\right) \;,
\end{align}
where we have used the fact $(-1)^{\hhhNR}$ commutes with the reservoir
Hamiltonian~\footnote{This would hold for a superconductor as well}
and, therefore, it is time independent.  This means that the Lindblad
equation controlling the time dependence of the \emph{reduced density
  matrix} of fermionic systems has the same form as for bosonic
ones, including its coefficients $C_{\alpha, \beta}$.

\subsection{Time evolution of fermionic operators}
The situation is different when considering correlation
  functions for  operators of the system, defined as
\beq 
\label{eq:Galphabeta} 
 G_{\beta,\al}(t) \equiv \tr_\fu \left( \fs^\dag_\beta(t) \
  \fs_\al \rho_\fu\right) = \tr_\fu \left( \fs^\dag_\beta
  \left(\fs_\al \rho_\fu \right)_t\right) \, ,  \eeq 
where $(\dots  )_t = e^{-i H_\fu t} (\dots ) e^{i H_\fu t}$ indicates
density-matrix-type time evolution as in Eq.\,(\ref{eq: tilde rho
  Hamiltonian dynamics}).
The standard 
QRT\cite{Breuer2002,Gardiner2000,carmichael1}  states that, within
 the BMA assumptions, 
the time evolution of operators
of the form 
$ \hhh\rhotilde_{\C}(t) = \tr_R(\C \rho_\fu)_t $ 
are governed
by the same Lindblad equation as $\hhh\rho(t)$,
namely (\ref{eq: Lindblad general}).
However, this theorem holds for operators $\C$ of the form
$\C=\hhh{\id}_\FRR \otimes \hhh{X}_\FS $.
As discussed above, due to the fermionic anticommutation rules, $\fs_\al$ does not have this form.
  However, it is possible to transcribe Eq.~\eqref{eq:Galphabeta}
into a form in which the 
standard QRT
can be applied
 by using the following scheme to keep track of
fermionic sign factors: 
\beqa 
\notag 
G_{\beta, \alpha} (t) & = \tr_\fu \left( \fs^\dag_\beta (-1)^N
  \left((-1)^N \fs_\al \rho_\fu \right)_t\right) 
\\
\notag & = \tr_\fu \left( \hhfs^\dag_\beta (-1)^\NS \left((-1)^\NS
    \hhfs_\al \rho_\fu \right)_t\right)
\\
\label{cfs} & = \tr_\FS \left( \hhh \fs^\dag_\beta (-1)^{\hhhNS} \ \tr_\FRR
  \left((-1)^\NS  \hhfs_\al \rho_\fu \right)_t\right)\! \,,
\eeqa
where $\NS=\hhh{\id}_R\otimes\hhhNS$ counts the number of particles in the system.
In the first line we exploited the fact that the operator for the
total number of particles in system and reservoir, $N = N_S + N_R$,
commutes with $H_\fu$.  In the second line we introduced the operator
\beq
\hhfs_\al \equiv (-1)^{\NR} \ \fs_\al = \hat{\id}_\FRR \otimes \hhh \fs_\al\;,
\eeq
which \emph{commutes} with the reservoir operators $\fr_\beta$.
Eq.~\eqref{cfs}
can be now cast in the form
\beq
G_{\beta,\alpha}(t) = \tr_\FS \left( \hhh \fs^\dag_\beta \hha_\alpha(t)\right) \, , 
\eeq
where we introduced  
\beqa
\label{hha} \hha_\al(t) & \equiv
 (-1)^{\hhhNS}\ \tr_\FRR\left((-1)^\NS\ \hhfs_\al \  \rho_\fu\right)_t 
\nonumber \\& =
 (-1)^{\hhhNS}\ \tr_\FRR\left(\left( \hat{\id}_\FRR \otimes (-1)^{\hhhNS}\ \hhh \fs_\al\right) \  \rho_\fu\right)_t 
\, ,  
\eeqa
which for $t=0$ reduces to $\hhh s_\alpha$ 
applied to the reduced system density
matrix \footnote{As for the bosonic case, $\rho$ can be previously
  have been time evolved up to a certain time $t_1$, which in steady
  state would be $t_1=\infty$.}:
 \beq 
\hha_\al(0)= \tr_\FRR\left( \hhfs_\al \rho_\fu\right) =
\hhh \fs_\al \hhh{\rho} \, .  
\eeq
Now, the operator multiplied to $\rho_\fu$
 in the last line in \eqref{hha} has the required form $\hat{\id}_\FRR \otimes \hat{X}_\FS$, so that,
 within the usual BMA assumptions,
 the QRT applies to the time dependence of the reservoir trace
in \eqref{hha}.
 Therefore, the time
evolution of $\hha_\alpha(t)$ 
yields
\beq \frac{d}{dt}
\hha_\al(t) = (-1)^{\hhhNS} \li\left(\tr_\FRR\left((-1)^\NS \hhfs_\al \rho_\fu \right)_t\right)
\equiv \underline{\li} \left( \hha_\al(t) \right) \; .  \eeq 
Here $\underline{\li}$ differs from Eq.\,(\ref{eq: Lindblad general}) by
having a minus sign in front of the $2 \hhh{J}_m \hhh{\rho}(t) \hhh{J}_m^\dagger$ term,
whenever $\hat{J}_m$ is a fermionic operator. For the quadratic system
discussed in Sec.~\ref{sec: Greens functions}, this leads to
Eq.\,(\ref{eq: rhotilde_dot}).

\bibliography{references}

\end{document}